%% file: main631-v202411_arxiv_v2.tex
\documentclass[twocolumn]{aastex631}
\usepackage{graphics, graphicx}
\usepackage{subfigure}
\usepackage{amsmath}
\usepackage{threeparttable}
\usepackage{booktabs}
\usepackage{tabularx}

\newcommand{\arcs}{\hbox{$^{\prime \prime}$}}
\newcommand{\arcm}{\hbox{$^{\prime}$}}
\newcommand{\degree}{\hbox{$^{\circ}$}}
\newcommand{\dmhost}{DM$_{\rm host}$}
\def\cxl#1 {{\textcolor{blue}{#1}}\ }
\def\bdw#1 {{\textcolor{green}{#1}}\ }

\def\Ha{H{$\rm{\alpha}$}}

\shorttitle{HOST GALAXY OF FRB\,20190520B}
\shortauthors{CHEN ET AL.}
\graphicspath{{./}{figures/}}

\begin{document}

\title{The Host Galaxy of FRB\,20190520B and Its Unique Ionized Gas Distribution}

\correspondingauthor{C.-W. Tsai; D. Li}
\email{cwtsai@nao.cas.cn; dili@tsinghua.edu.cn}

\author[0000-0001-5738-9625]{Xiang-Lei Chen}
\affiliation{National Astronomical Observatories, Chinese Academy of Sciences, 20A Datun Road, Beijing 100101, China}

\author[0000-0002-9390-9672]{Chao-Wei Tsai}
\affiliation{National Astronomical Observatories, Chinese Academy of Sciences, 20A Datun Road, Beijing 100101, China}
\affiliation{Institute for Frontiers in Astronomy and Astrophysics, Beijing Normal University,  Beijing 102206, China}
\affiliation{University of Chinese Academy of Sciences, Beijing 100049, China}

\author[0000-0003-2686-9241]{Daniel Stern}
\affiliation{Jet Propulsion Laboratory, California Institute of Technology, 4800 Oak Grove Drive, Pasadena, CA 91109, USA}

\author[0000-0003-3875-9568]{Christopher D. Bochenek}
\affiliation{Division of Physics, Math, and Astronomy, California Institute of Technology, 1200 E California Blvd., Pasadena, CA 91125, USA}

\author[0000-0002-2878-1502]{Shami Chatterjee}
\affiliation{Department of Astronomy, Cornell University, Ithaca, New York 14853, USA}
\affiliation{Cornell Center for Astrophysics and Planetary Science, and Department of Astronomy, Cornell University, Ithaca, NY, USA}

\author[0000-0002-4119-9963]{Casey Law}
\affiliation{Cahill Center for Astronomy and Astrophysics, MC 249-17 California Institute of Technology, Pasadena CA 91125, USA}
\affiliation{Owens Valley Radio Observatory, California Institute of Technology, Big Pine CA 93513, USA}

\author[0000-0003-3010-7661]{Di Li}
\affiliation{Department of Astronomy, Tsinghua University, Beijing 100084, China}
\affiliation{National Astronomical Observatories, Chinese Academy of Sciences, 20A Datun Road, Beijing 100101, China}
\affiliation{Zhejiang Lab, Hangzhou, Zhejiang 311121, China}

\author[0000-0001-6651-7799]{Chenhui Niu}
\affiliation{Institute of Astrophysics, Central China Normal University, Wuhan 430079, China}

\author[0000-0001-5322-5076]{Yuu Niino}
\affiliation{Kiso Observatory, Institute of Astronomy, Graduate School of Science, The University of Tokyo, 10762-30 Mitake, Kiso-machi, Kiso-gun, Nagano 397-0101, Japan}

\author[0000-0002-0475-7479]{Yi Feng}
\affiliation{Research Center for Astronomical Computing, Zhejiang Laboratory, Hangzhou 311100, China}
\affil{Institute for Astronomy, School of Physics, Zhejiang University, Hangzhou 310027, China}

\author[0000-0002-3386-7159]{Pei Wang}
\affiliation{National Astronomical Observatories, Chinese Academy of Sciences, 20A Datun Road, Beijing 100101, China}
\affiliation{Institute for Frontiers in Astronomy and Astrophysics, Beijing Normal University,  Beijing 102206, China}

\author[0000-0002-9508-3667]{Roberto J. Assef}
\affiliation{Instituto de Estudios Astrofísicos, Facultad de Ingeniería y Ciencias, Universidad Diego Portales, Ej´ercito Libertador 441,
8370191, Santiago, Chile}

\author[0000-0003-4007-5771]{Guodong Li}
\affiliation{Kavli Institute for Astronomy and Astrophysics, Peking University, Beijing 100871, People’s Republic of China}

\author[0000-0002-4528-7637]{Sean E. Lake}
\affiliation{National Astronomical Observatories, Chinese Academy of Sciences, 20A Datun Road, Beijing 100101, China}

\author[0000-0002-1583-8514]{Gan Luo}
\affiliation{Institut de Radioastronomie Millimetrique, 300 rue de la Piscine, 38400, Saint-Martin d’Hères, France}

\author[0000-0002-9137-7019]{Mai Liao}
\affiliation{National Astronomical Observatories, Chinese Academy of Sciences, 20A Datun Road, Beijing 100101, China}
\affiliation{Universidad Diego Portales, Av Republica 180, Santiago, Reg\'{i}\'{o}n  Metropolitana, Chile}

\begin{abstract}

The properties of host galaxies associated with Fast Radio Bursts (FRBs) provide critical information for inferring the progenitors and radiation mechanisms of these bursts.
We report on the host galaxy of the repeating FRB 20190520B, a dwarf galaxy at the spectroscopic redshift $z=0.241$ with a stellar mass of $(6.2 \pm 0.8) \times 10^8 \ M_{\odot}$.
The emission line ratios suggest that the ionized gas is powered by star formation.
The total \Ha-traced star formation rate (SFR) is $0.70 \pm 0.01 \ {M_{\odot} ~ \rm yr^{-1}}$, and the metallicity is $\rm 12+log_{10} ([O/H]) \geq 7.4 \pm 0.1$.
The specific star formation rate (sSFR) is $\rm log \ sSFR/yr^{-1} = -9.0 \pm 0.1$, higher than the upper limit of $-9.4$ observed in nearby dwarf galaxies.
The dispersion measure contribution from the host galaxy is estimated to be $\rm DM_{host} \approx 950 \pm 220 \ pc \ cm^{-3}$, based on the \Ha\ emission.
The FRB and the associated persistent radio source are located at the \Ha\ emission peak, offset by $\sim 1.4$\arcsec\ (5.5 kpc) in projection from the stellar continuum.
At this position, the lower limit of $\rm \log \ sSFR/yr^{-1}$ is $-8.5 \pm 0.1$, more than three times the galaxy's total sSFR.
The \Ha\ velocity difference between the stellar continuum and the offset gas is $39.6 \pm 0.4$ km s$^{-1}$, which is sufficient to draw conclusions about the nature of the offset.

\end{abstract}


\section{Introduction} \label{sec:intro}

Fast Radio Bursts (FRBs) are radio transients with an unknown origin and radiation mechanism, with durations ranging from milliseconds to seconds \citep{Zhang2023RvMP...95c5005Z}. 
Since their discovery \citep{Lorimer2007Sci...318..777L}, more than eight hundred FRBs have been detected\footnote{\url{https://www.wis-tns.org}}$^,$\footnote{\url{https://blinkverse.zero2x.org}} \citep{blinkverse}, with energy releases during these short events reaching up to $\sim 10^{39}$ erg \citep{Petroff2022A&ARv..30....2P}.
This has sparked significant interest in studying the formation mechanism(s) and energy source(s) of FRBs, and led to the development of a multitude of theoretical models.\footnote{\url{https://frbtheorycat.org}}. 
Most FRBs are believed to have extragalactic origins, except for FRB 20200428 \citep[SGR 1935+2154,][]{SGR19352020Natur.587...54C}.

During propagation, the plasma in the Interstellar Medium (ISM) and Intergalactic Medium (IGM) causes pulse dispersion in FRBs, detected as a frequency-dependent delay in the arrival times of the pulses. 
The extent of the time delay depends on the electron density along the line of sight.
Therefore, the distribution of ionized baryons between the FRB and the observer can be probed through the Dispersion Measure, DM$ \equiv \int_0^s { n_e ds } $, where $n_e$ is the electron density along the line of sight, and $ds$ is the line-of-sight distance element. 
The integrated effect shows a strong correlation with the FRB's redshift. 
The so-called Macquart relation \citep{MacquartRelation2020Natur.581..391M} offers a valuable estimate of the FRB's distance based on its observed DM, and subsequently, its luminosity. 
This relation is a key assumption for employing FRBs as cosmological probes, however the contribution from the Milky Way and, most importantly, the FRBs host galaxy add significant uncertainty \citep{RaviWP2019BAAS...51c.420R}.

FRBs can be phenomenologically divided into repeaters and one-off sources, depending on the number of observed pulses. 
Historically, the high DM measured for the first discovered repeater, FRB 20121102A, significantly exceeded the expected contribution from the Milky Way's ISM, implying an extragalactic origin \citep{Spitler2014ApJ...790..101S, Spitler2016Natur.531..202S} which was later confirmed by optical followup observations. 
The properties of the host galaxy of FRB 20121102A are similar to those of the host galaxies of long gamma-ray bursts (LGRBs) and superluminous supernovae (SLSNe), characterized by low stellar masses, low metallicities, and high star formation rates, suggesting a potential relationship between these sources and FRBs \citep{Zhang2023RvMP...95c5005Z}.  
Since the discovery of FRB 20121102A, the host galaxies of more than ninety FRBs have been identified.

FRB 20201124A, another active repeater, is located in a low-stellar-density interarm region of a Milky Way-sized barred spiral galaxy at $z=0.0979$ \citep{Ravi201124A2022MNRAS.513..982R, Xu2011242022Natur.609..685X}.
Later, radio continuum observations identified this location as an obscured star-forming region \citep{Dong2024ApJ...961...44D}. 
The high degree of circular polarization observed in FRB 20240114A may suggest a complex, dynamically evolving magnetized local environment.
In contrast, the repeating FRB 20220912A \citep{220912Ravi2023ApJ...949L...3R} has a high burst rate of up to 390 hr$^{-1}$ and a low rotation measure (RM) of $ -0.08 \pm 5.39 \ \rm rad \ m^{-2}$ in the \textit{L}-band \citep{220912ZYK2023ApJ...955..142Z}, suggesting a clean environment. 
Optical observations reveal the disk-like morphology of its host galaxy at a redshift of $z=0.0771$. 
The significant diversity in host galaxy properties and inferred stellar populations not only provides clues about the origin of FRBs but also suggests the possibility of multiple formation mechanisms and progenitors for FRBs.

The repeating FRB 20190520B \citep{Niu190520-2022Natur.606..873N} was discovered by the Five-hundred-meter Aperture Spherical Radio Telescope \citep[FAST,][]{NanFAST2011IJMPD..20..989N} in the Commensal Radio Astronomy FAST Survey \citep[CRAFTS,][]{LiCRAFTS2018}, and was then localized using the \textit{realfast} system \citep{realfast} of the Very Large Array (VLA). 
An unresolved compact persistent radio source (PRS) was detected by the VLA at a distance of 0.16 arcsec from the FRB location, marking it as the second FRB associated with a PRS, following FRB 20121102A.  
Subsequent observations with the European VLBI Network (EVN) further constrained the separation to $\leq 0.02^{\prime\prime}$ \citep{190520PRS_VLBI}.  
The faint host galaxy was identified through spectroscopic observations at the VLA positions of the FRB and the PRS, where distinct emission lines were detected.
FRB 20190520B exhibits an observed high DM of $\sim 1205 \pm 4$ pc cm$^{-3}$.
This includes a foreground DM contribution of $\sim 110 \pm 20$ pc cm$^{-3}$ from the Milky Way, estimated using the disk models of NE2001 \citep[$\sim 60$ pc cm$^{-3}$, ][]{NE2001-12002astro.ph..7156C, NE2001-22003astro.ph..1598C} and YMW \citep[$\sim 50$ pc cm$^{-3}$, ][]{YMW2017ApJ...835...29Y}, as well as the halo model \citep[$\sim 70$ pc cm$^{-3}$, ][]{DMhalo2020ApJ...888..105Y}.

The estimated extragalactic DM significantly exceeds the value predicted by the Macquart relation for a source at redshift $z=0.241$ \citep[Figure~3 in][]{Niu190520-2022Natur.606..873N}, with the host galaxy contributing an estimated $\rm DM_{host} \sim 900$ pc cm$^{-3}$.
Additionally, \cite{foreground2023ApJ...954L...7L} conducted a spectroscopic survey to search for foreground galaxy clusters along the sightline of FRB 20190520B, identifying two clusters in its path. Their analysis suggests that these foreground clusters could contribute an additional $ 450 - 640 $ pc cm$^{-3}$ to the extragalactic DM.

Follow-up observations by \cite{2022SciBu..67.2398F} using the FAST detected bursts with circular polarization.  
Follow-up observations by \cite{FengsigRM2022Sci...375.1266F} using the Robert C. Byrd Green Bank Telescope (GBT) detected bursts with high RM (averaging 2759 rad m$^{-2}$) and a large RM scatter ($\sigma_{\rm{RM}} = 218.9$ rad m$^{-2}$), suggesting that $\sigma_{\rm{RM}}$ traces the complexity of the magneto-ionic environment around the FRB.
Later RM reversal was detected in \cite{reshma23}, which is consistent with the bursts passing through the stellar wind of a binary companion of the FRB source.  
\cite{OckerHost2022ApJ...931...87O} analyzed the large dispersion and pulse scattering of FRB 20190520B, attributing them to the host galaxy.
They find a mean scattering time of $\tau = 10.9 \pm 1.5$ ms at 1.41 GHz, implying that the distance ($l_X$) between the FRB emitter and the dominant scattering material is less than 100 pc. 
\cite{OckerScatter2023MNRAS.519..821O} demonstrated that dynamic and inhomogeneous plasma in the circum-FRB medium causes scattering variability in FRB 20190520B, similar to that of the Crab pulsar. 
All of these DM and RM analyses point to a rich and complex ionized environment surrounding FRB 20190520B.

In this work, we report on optical imaging and spectroscopy of the host galaxy of FRB 20190520B to investigate the environment of the FRB. 
This paper is organized as follows: 
Section~\ref{sect:obs} provides details of the observations. 
Section~\ref{sect:data} introduces the data reductions.
Section~\ref{sect:analy} describes the data analysis, and compares the star-forming properties with those of other FRB host galaxies.
Section~\ref{sect:ionizedgas} presents the inferred distribution of ionized gas in the host galaxy. 
Section~\ref{sect:discussion} discusses the DM contribution of the ionized gas in the host galaxy, and the PRS associated with FRB 20190520B. 
Finally, Section~\ref{sect:sum} summarizes our results.
We adopt a flat $\Lambda$CDM cosmology with $H_0 = 67.66$ km s$^{-1}$ Mpc$^{-1}$ and $\Omega_{\rm m}=0.310$ \citep{Planck2020A&A...641A...6P}.
We also adopt a solar abundance of $\rm 12+log_{10} ([O/H]) = 8.69 \pm 0.04$ \citep{solar3-2021A&A...653A.141A}.
At the redshift $z=0.241$ of the host galaxy of FRB 20190520B, 1$^{\prime \prime}$ corresponds to 3.8 kpc.

\section{Observations} \label{sect:obs}

The active repeating FRB 20190520B, discovered by FAST, was localized by VLA at (R.A., DEC.)[J2000]$ = (\rm 16h02m04.272s,\ -11^{\circ}17^{\prime}17.32^{\prime \prime})$ with 1$\sigma$ positional uncertainties of 0.10$^{\prime \prime}$ and 0.08$^{\prime \prime}$, respectively \citep{Niu190520-2022Natur.606..873N}. 
The coordinates of PRS used in this work was localized by VLA at (R.A., DEC.)[J2000]$ = (\rm 16h02m04.261s,\ -11^{\circ}17^{\prime}17.350^{\prime \prime})$, approximately $0.165^{\prime \prime}$ to the west of FRB 20190520B, with 1$\sigma$ uncertainties of 0.10$^{\prime \prime}$ and 0.05$^{\prime \prime}$, respectively.
We also note that the ENV-localized coordinates of PRS are (R.A., DEC.)[J2000]$ = (\rm 16h02m04.2611s,\ -11^{\circ}17^{\prime}17.366^{\prime \prime})$, with uncertainties of 6.5 mas and 3.6 mas, respectively, which fall entirely within the VLA positional uncertainties of the PRS.

\subsection{The Optical and Near-Infrared Images with CFHT, GTC and Subaru} \label{subsect:img}

The optical \textit{R\arcm}- and \textit{GRI}-band images of the host galaxy of FRB 20190520B were obtained from the CFHT/MegaCAM archival data taken on April 10, 2013, and March 28, 2017, respectively. 
The \textit{R\arcm}-band image was shown in the panel a of Fig.~2 in \cite{Niu190520-2022Natur.606..873N} and is also used in this work.

\begin{figure}[htb]
    \centering
    \includegraphics[width=0.9\linewidth]{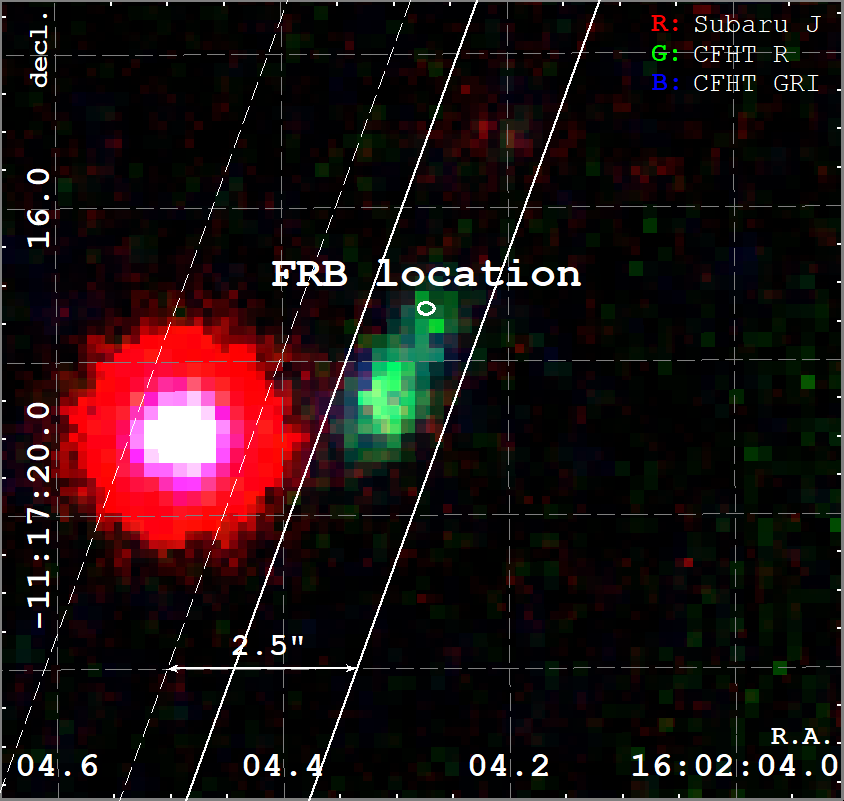}
    \caption{
    Pseudo-RGB image of the host galaxy of FRB 20190520B, generated using the Subaru \textit{J}-band (Red) and CFHT \textit{R\arcm}-band (Green) and \textit{GRI}-band (Blue) images. 
    The white ellipse indicates the location of the FRB along with the reported uncertainties from \cite{Niu190520-2022Natur.606..873N}. 
    The white solid lines represent the Keck/LRIS long slit, which was at a position angle of PA$=160^{\circ}$, centered on the nearby M-type star (dashed white lines), and then shifted westward by $2.5^{\prime \prime}$.
    }
    \label{fig:rgb}
\end{figure}

A near-infrared \textit{J}-band image was acquired on August 5, 2020, using the Subaru/MOIRCS in target-of-opportunity mode. 
Figure~\ref{fig:rgb} presents a pseudo-RGB image centered on the host galaxy, generated from these three images. 
The red source to the east of the host galaxy is an M-type star.
 
We also obtained broad-band images using OSIRIS$+$ on the Gran Telescopio Canarias (GTC) under program ID GTCMULTIPLE1A-23ACNT (PI: C.-W. Tsai). 
The images in the \textit{r}- and \textit{i}-bands were captured on July 6, 2023 with dithering and the observations were conducted under seeing conditions of $\sim1.1^{\prime \prime}$ and without moonlight.

\subsection{Spectroscopic Observations with Palomar/DBSP} \label{subsect:spec}

A detailed description of the Palomar 200-inch telescope (P200) observations are discussed in \citet{Niu190520-2022Natur.606..873N}. 
We briefly review them here.

Prior to obtaining the archival optical images from CFHT, the long-slit spectrum was acquired using the Double Spectrograph (DBSP) on the P200 on July 24, 2020. 
The selected grating had 316 lines mm$^{-1}$ and was blazed at 7500 \AA. 
The slit width was set to $1^{\prime \prime}$, and the grating angle was 24.63$^{\circ}$. 
The observations were conducted under sub-arcsecond seeing conditions. 
A total of $2 \times 900$ s of observations were obtained. 
The redshift $z = 0.241 \pm 0.001$ of the host galaxy was determined based on the clearly detected narrow emission lines of [OIII]$\lambda$5007 and H$\alpha$ \citep[][Extended Data, Figure~3]{Niu190520-2022Natur.606..873N}. 
This redshift was later confirmed with the Keck/LRIS spectroscopic data described in the next section.

\subsection{Spectroscopic Observations from Keck/LRIS} \label{subsubsect:Keckobs}

The P200/DBSP observation was initially designed based on the localization of the PRS without identifying the host galaxy, covering only approximately 4650–7250 \AA\ in the rest frame with a low signal-to-noise ratio.
To address this, we conducted a subsequent spectroscopic observation using the Low Resolution Imaging Spectrometer (LRIS) on the Keck I telescope on August 25, 2020, under seeing conditions of approximately 1.1\arcs.
Three spectra, each with an exposure time of 900 s, were obtained using a long slit with a width of 1.5\arcs\ and a position angle of 160\degree \space aligned with the elongated direction of the host galaxy (see Fig.~\ref{fig:rgb}).
For the blue arm of LRIS we used the 400/3400 grism, while for the red arm we used the 400/8500 grating.
Together, this instrument configuration covers the wavelength range from 3100 \AA \space to 10280 \AA. 
The standard star BD$+$33 2642 was used for flux calibration. 
Figure~\ref{fig:keck-spec} displays the combined spectrum derived from the three exposures.

\section{Data Reductions} \label{sect:data}

\subsection{GTC Images}

We obtained deep \textit{r}- and \textit{i}-band images of the host galaxy and its environment using the broadband image mode of the GTC/OSIRIS$+$.
The raw data were first bias-subtracted, flat-fielded, cleaned of cosmic rays, and combined using the \textit{ccdproc} \citep{ccdprocmatt_craig_2017_1069648} and \textit{DrizzlePac} \citep{drizzle2021AAS...23821602H} packages, following standard techniques. 
The \textit{r}-band image was produced by co-adding all four single-exposure images with 150 sec exposure time for each frame. 
One of the four \textit{i}-band images is significantly affected by stellar spikes from a nearby bright star ($i=8.5$ mag, $58.84^{\prime \prime}$ to the south-west).
The remaining three exposures were combined to generate the final \textit{i}-band image.
A 600 sec total exposure was obtained for the final \textit{r}-band image, and 540 sec was obtained for the final \textit{i}-band image.
There is significant contamination from the nearby star east of the galaxy, making careful sky subtraction crucial for accurate target galaxy measurements. 
We modeled and subtracted the complex 2D background of the total image primarily using the \textit{photutils} package, following instructions in the Cross-Instrument section of the JWST Data Analysis Tool Notebooks\footnote{\url{https://spacetelescope.github.io/jdat_notebooks/notebooks/cross_instrument/background_estimation_imaging/Imaging_Sky_Background_Estimation.html}}. 
Photometry was performed on the sky-subtracted image using an elliptical aperture.

\subsection{Keck Spectrum}

\begin{figure*}[ht!]
    \centering
    \includegraphics[width=0.9\textwidth]{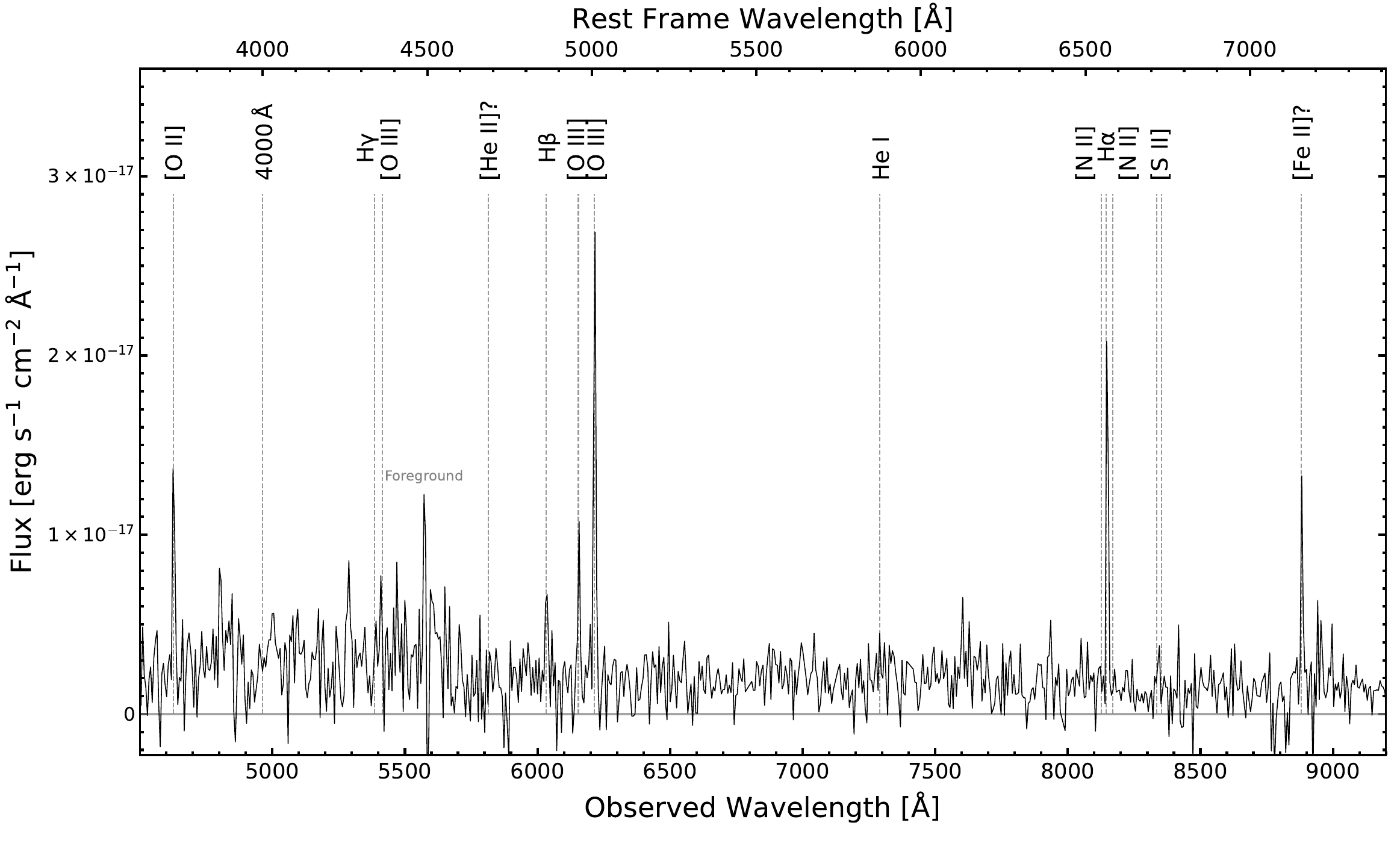}
    \caption{Keck/LRIS spectrum of the host galaxy of FRB 20190520B
    at $z = 0.241 \pm 0.001$.
    The lower and upper abscissas represent the observed and rest-frame wavelengths, respectively. 
    The false emission line caused by foreground contamination is indicated as ``Foreground''. 
    }
    \label{fig:keck-spec}
\end{figure*}

We processed the Keck spectroscopic data with IRAF using standard techniques.
Figure~\ref{fig:keck-spec} shows the 1D spectrum obtained by combining the blue and red side spectra. 
In addition to the H$\alpha$, H$\beta$, and [O III] emission lines observed by the P200, blended [O II]$\lambda$3728 emission lines are also detected. 
The host galaxy redshift is determined to be $z = 0.241\pm0.001$ based on these observed lines. 
Figure~\ref{fig:keck-spec} also displays blended [Fe II]$\lambda$7151/56 emission lines. 
Typically, the presence of [Fe II]$\lambda$7151/56 emission indicates ionized gas excited by supernovae; however, no other supernova emission features, such as [Fe III]$\lambda 4658$, [Fe II]$\lambda 5195$, or [Ca II]$\lambda 7291/7323$ \citep{SLSNe2013Natur.502..346N, IaSpec2018MNRAS.477.3567M}, are observed in the spectrum. 
We suggest that the observed line may be a spurious emission resulting from a cosmic ray.

\section{Data Analysis} \label{sect:analy}

\subsection{Spectrum Analysis} \label{subsect:spec_ana}

The emission lines in the co-added spectrum of the host galaxy of FRB 20190520B were fitted using single-Gaussian profiles. 
The line fluxes, after correcting for Galactic extinction \citep{SFD1998ApJ...500..525S}, are listed in Table~\ref{tab:em-lines}.
The upper limits for the H$\gamma$ and [O III]$\lambda$4363 emission lines were estimated by assuming that their line widths are identical to those of H$\beta$ and [O III]$\lambda$5007, respectively. 
The intrinsic extinction is estimated to be $A_{\rm{V}} = 1.24 \pm 0.31 $ based on the line flux ratio between H$\alpha$ and H$\beta$, after correcting for Galactic foreground extinction and assuming an intrinsic case-B line ratio of 2.86. 
This value is characteristic of similar starburst dwarf galaxies \citep{Griffith2011ApJ...736L..22G}.

\begin{table}[htbp]
\centering
    \caption{Emission line properties of the host galaxy of FRB 20190520B at the rest frame.
    }
    \medskip
    \resizebox{\linewidth}{!}{
    \begin{threeparttable}
    \begin{tabular}{lcc}
    \hline
    \hline
    \noalign{\smallskip}
         Line                    & F$_{\lambda}$                              & EW  \\
                                 & ($10 ^{-16}$ erg cm$^{-2}$ s$^{-1}$)       & (\AA) \\
         \noalign{\smallskip}
         \hline
         \noalign{\smallskip}
         [O II] $\lambda$ 3728  & $17.0 \pm 2$    & $50 \pm 10$ \\
         \noalign{\smallskip}
         H $\gamma^{\ast}$             & $<2.1$ & $<6$ \\
         \noalign{\smallskip}
         [O III] $\lambda$ 4363$^{\ast}$ & $<1.8$ & $<6$ \\
         \noalign{\smallskip}
         H $\beta$              & $2.5 \pm 0.3 $ & $12^{+2}_{-2}$ \\
         \noalign{\smallskip}
         [O III] $\lambda$ 4959 & $4.8 \pm 0.3 $ & $25 \pm 3 $ \\
         \noalign{\smallskip}
         [O III] $\lambda$ 5007 & $16.2 \pm 0.4 $ & $110^{+20}_{-10}$ \\
         \noalign{\smallskip}
         He I $\lambda$ 5875    & $0.8^{+0.5}_{-0.3} $ & $6^{+5}_{-3}$ \\
         \noalign{\smallskip}
         [N II] $\lambda$ 6548  & $0.3 \pm 0.1 $ & $5^{+1}_{-2}$ \\
         \noalign{\smallskip}
         H$\alpha$              & $7.0 \pm 0.1 $ & $116^{+8}_{-7}$ \\
         \noalign{\smallskip}
         [N II] $\lambda$ 6583  & $0.2 \pm 0.1 $ & $4^{+2}_{-1}$ \\
         \noalign{\smallskip}
         [S II] $\lambda$ 6716  & $0.3 \pm 0.1 $ & $4^{+2}_{-1}$ \\
         \noalign{\smallskip}
         [S II] $\lambda$ 6731  & $0.19 \pm 0.1 $ & $3^{+2}_{-1}$ \\
         \noalign{\smallskip}
         \hline
    \end{tabular}
    \medskip
    \begin{tablenotes}
    \footnotesize 
    \item[$\ast$] The upper limits for the H$\gamma$ and [O III]$\lambda$4363 emission lines are reported at a significance level of $3\sigma$.
    \end{tablenotes}
\end{threeparttable}
}
    \label{tab:em-lines}
\end{table}

\begin{figure}[ht]
    \centering
    \includegraphics[width=\linewidth]{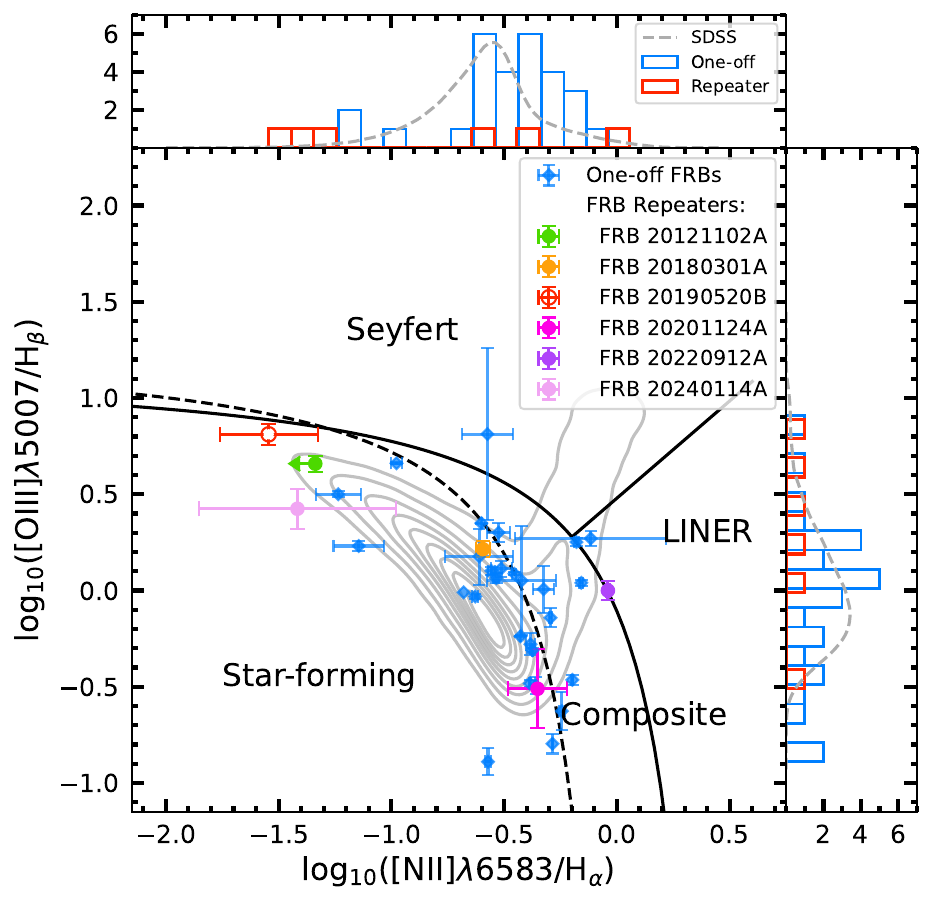}
    \caption{
    BPT classification diagram for the host galaxies of FRBs.
    The host galaxies of repeaters are labeled with their names.
    The property of FRB 20190520B's host galaxy is shown by the unfilled circle.
    The distributions of repeating and one-off FRB host galaxies are shown as red and blue histograms, respectively.
    Contours represent SDSS DR17 galaxies with significant emission lines ($>5\sigma$), and their scaled kernel density estimate (KDE) is shown in gray.
    The black lines mark the boundaries between star-forming galaxies and composite sources \citep{Kauffmann2003MNRAS.346.1055K}, composite sources and LINERs \citep{Kewley2006MNRAS.372..961K}, and LINERs and Seyfert galaxies \citep{Fernandes2010MNRAS.403.1036C}.
    }
    \label{fig:bpt}
\end{figure}

Figure~\ref{fig:bpt} shows the Baldwin, Phillips, and Terlevich (BPT) diagram \citep{BPT1981PASP...93....5B} for the host galaxy of FRBs, along with SDSS DR17 galaxies at redshifts $0.02 < z < 0.4$ \citep{SDSSDR172022ApJS..259...35A} for comparison. 
The line ratios for the host galaxies of the repeating FRBs are highlighted in distinct colors. 
The BPT analysis of the FRB 20190520B host galaxy indicates that the observed emission lines are consistent with star formation. 
This is similar to the metal-poor star-forming dwarf host galaxies of FRB 20121102A and FRB 20240114A.

We derived an SFR of $0.70 \pm 0.01 \ M_{\odot} \, \mathrm{yr}^{-1}$ from the extinction-corrected H$\alpha$ luminosity $L_{\mathrm{H}\alpha} = (1.3 \pm 0.1) \times 10^{41} \, \mathrm{erg \, s}^{-1}$ \citep{MurphyEJ2011, HaoCN2011, SFR2012ARA&A..50..531K}. 
The stellar mass is estimated using the \textit{J}-band magnitude and the ($r-i$) color following the method in \cite{Bell2003ApJS..149..289B}.
We derived a stellar mass of $(6.2 \pm 0.8) \times 10^8 \ M_{\odot}$ (log M$_* = (8.9 \pm 0.1) \ M_{\odot}$), which lies between the values reported in \cite{Niu190520-2022Natur.606..873N} and \cite{Gordon2023hostgalaxies}.
The estimated specific star formation rate (sSFR) is $\rm{log \, sSFR/yr^{-1}} = -9.0 \pm 0.1 $, which is higher than the upper limit of $\rm{log \, sSFR/yr^{-1} \simeq -9.4} $ observed in nearby dwarf galaxies \citep{2013AJ....146...46K}.

We estimate the metallicity of the host galaxy of FRB 20190520B using the T$_e$-method \citep{Aller1984ASSL..112.....A,Izotov2006A&A...448..955I}. 
Since the estimated flux of [O III]$\lambda$4363 is an upper limit, we provide an upper limit for the electron temperature and a lower limit for the metallicity. 
The electron temperature derived from [O III] is $T_e[{\rm O \ III}] \leq (3.2 \pm 1.6) \times 10^4 \ {\rm K}$. 
The corresponding metallicity is $12 + {\rm log_{10}([O/H])} \geq (7.4 \pm 0.1)$.
These derived parameters are listed in Table~\ref{tab:gal}.

\begin{table}[htbp]
    \centering
    \caption{Properties of the host galaxy of FRB 20190520B.}
    \medskip
    \begin{threeparttable}
    \begin{tabular}{lc}
    \hline
    \hline
    \noalign{\smallskip}
         Redshift & 0.241 $\pm$ 0.001 \\
         \noalign{\smallskip}
         \textit{r} (AB mag) & 22.41 $\pm$ 0.04 \\
         \noalign{\smallskip}
         \textit{i} (AB mag) & 22.34 $\pm$ 0.07 \\
         \noalign{\smallskip}
         \textit{R\arcm} (AB mag) & 22.73 $\pm$ 0.03 \\
         \noalign{\smallskip}
         \textit{J}~~(AB mag) & 22.07 $\pm$ 0.14 \\
         \noalign{\smallskip}
         Stellar Mass$^{\dagger}$ ($\rm log_{10}\mathcal{M}_{\odot}$) & $8.8 \pm 0.1 $ \\
         \noalign{\smallskip}
         $T_e[{\rm O\ III}]$ ($10^4$ K) & $ 3.2 \pm 1.6$  \\
         \noalign{\smallskip}
         $ 12+ {\rm log_{10} \ ([O/H])} $ & $\geq 7.4 \pm 0.1$ \\
         \noalign{\smallskip}
         $L_{\rm{H\alpha}}$ (erg s$^{-1}$) & $1.3 \pm 0.1 \times 10^{41} $ \\
         \noalign{\smallskip}
         SFR(H${\alpha}$) ($\mathcal{M}_{\odot} \ {\rm yr}^{-1}$)  & $ 0.70 \pm 0.01 $ \\
         \noalign{\smallskip}
         log sSFR (yr$^{-1}$) & $-9.0 \pm 0.1$ \\
         \noalign{\smallskip}
         $ S({\rm H \alpha})_s $ (Rayleigh)$^{\ast}$ & $460 \pm 7$ \\
         \noalign{\smallskip}
         DM(H$\alpha$)$_s$ (pc cm $^{-3}$)$^{\diamond} $ & $950\pm 220$ \\
         \noalign{\smallskip}
         \hline
    \end{tabular}
    \medskip
    \begin{tablenotes}
    \footnotesize
    \item[$\dagger$] Based on the \textit{J}-band magnitude.
    \item[$\ast$] Assuming a galaxy size of $r=0.45 ^{\prime \prime} $. $ S({\rm H \alpha})_s $ is equal to $2.61 \pm 0.04 \times 10^{-15}$ erg cm$^{-2}$ s$^{-1}$ arcsec$^{-2}$.
    \item[$\diamond$] Assuming $l=5$ kpc, $f=0.1$ and $\zeta=\epsilon^2=1$.
    \end{tablenotes}
    \end{threeparttable}
    \label{tab:gal}
\end{table}

\subsection{Stellar Mass and SED Fitting} \label{subsect:SED}

We model the optical to near-infrared (NIR) spectral energy distribution (SED) to decompose the stellar components in the host galaxy of FRB 20190520B following the methodology described by \citet{2015ApJ...804...27A}.
The magnitudes in the \textit{r}- and \textit{i}-bands are the photometric results from GTC broad-band images, while the CFHT \textit{GRI}- and \textit{R}-band magnitudes are measured using archived data from CFHT. 
The \textit{J}-band magnitude is adopted from \cite{Niu190520-2022Natur.606..873N}.
Additionally, we used the \textit{g}- and \textit{z}-band magnitudes from \cite{Gordon2023hostgalaxies} in the fittings. 
The \textit{r}-band magnitude from \cite{Gordon2023hostgalaxies} was not used in our analysis, as the photometric error using the GTC data is smaller. 
The upper limit for the \textit{u}-band from \cite{Gordon2023hostgalaxies} is also excluded from the SED fitting. 
Thus, data used in the SED fitting spans from observed 3800 \AA\ to 13000 \AA.

The SED of a galaxy is represented as a linear combination of four empirical spectral templates \citep{2010ApJ...713..970A}. 
Three of these templates correspond to galaxy SEDs: ``E" for an elliptical galaxy with an old stellar population, ``Sbc" for a spiral galaxy with intermediate star-forming activities, and ``Im" for an irregular Magellanic galaxy with starburst activities. 
The fourth template corresponds to an unreddened AGN. 
We fit the SED using the Markov Chain Monte Carlo (MCMC) implementation described by \citet{emcmc2013PASP..125..306F} through the public Python package \textit{emcee}\footnote{\url{https://github.com/dfm/emcee}}. 
The flux of each band was corrected for extinction from dust in the Milky Way prior to the modeling process \citep[$A_{\rm V} = 0.77 \ \text{mag}$;][]{SFD1998ApJ...500..525S,2007ApJ...663..320F}. 
We use uninformative priors for all parameters, enforcing them to be non-negative.
The median of the marginalized distribution of each parameter is taken as its best-fit value, with the 1$\sigma$ uncertainties corresponding to the 16th and 84th percentiles of the distributions.

\begin{figure}[htb]
    \centering
    \includegraphics[width=\linewidth]{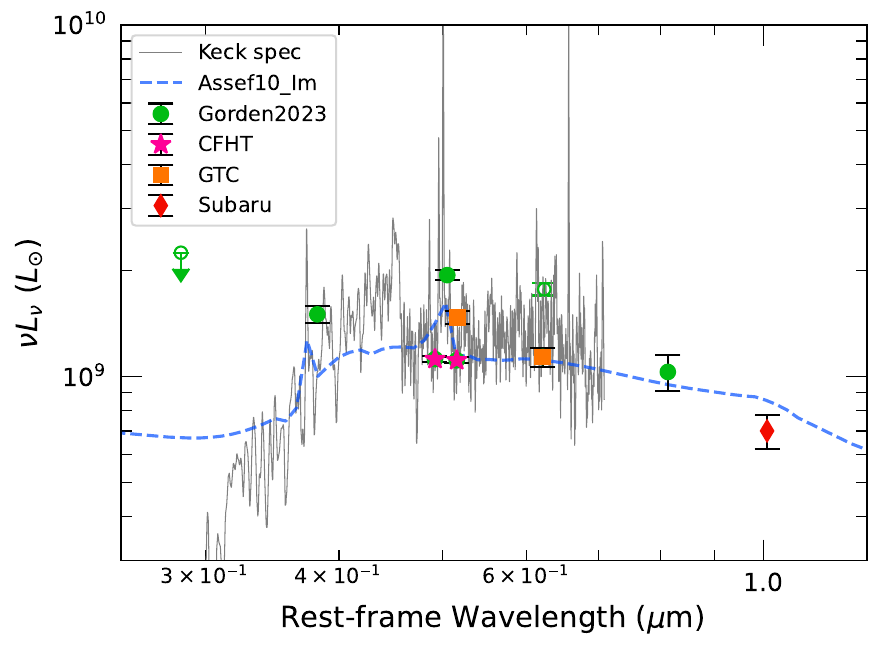}
    \caption{
    Best-fit SED template model of the host galaxy of FRB 20190520B. 
    All photometry described in Sect.~\ref{subsect:SED}, except for the \textit{u}-band upper limit, was used in the fitting. 
    The SED modeling employs the template of an irregular galaxy from \citet{2010ApJ...713..970A}, as described in Section \ref{subsect:SED}.}
    \label{fig:sed}
\end{figure}

Figure~\ref{fig:sed} shows the best-fit SED model of the host galaxy of FRB 20190520B. 
Preliminary testing found that the ``Im" template dominates its optical-NIR SED, while the other components (``E", ``Sbc", and ``AGN") were negligible, suggesting that the host galaxy is likely a star-forming system. 
We estimate its optical luminosity to be $L_{\rm optical} \sim 1.1\times 10^{9} \ L_{\odot}$ by integrating from 0.382 to 1.010 $\mu$m in the rest frame, using power-law interpolation on the best-fit SED model \citep{2023ApJ...958..162L}. 
Since the galaxy was not detected at mid-infrared wavelengths by \textit{WISE} \citep{WISE2010AJ....140.1868W}, we cannot estimate its infrared luminosity. 
The optical luminosity is consistent with the value derived by \citet{Gordon2023hostgalaxies}.
Using the synthesized magnitudes and the correlations between \textit{ugriz} colors and mass-to-light ratios reported by \citet{Bell2003ApJS..149..289B}, the estimated stellar mass from the SED fitting is $\rm log_{10} \ M_*/ \it M_{\odot} \rm = 8.9 \pm 0.1 $.
The stellar mass estimated by \cite{Gordon2023hostgalaxies} using SED fitting is $\rm log_{10} \ M_*/ \it M_{\odot} \rm = 9.08^{+0.08}_{-0.09} $, which is slightly higher than our results. 

\begin{figure}[ht]
    \centering
    \subfigure{
    \begin{minipage}[h]{\linewidth}
        \includegraphics[width=0.99\linewidth]{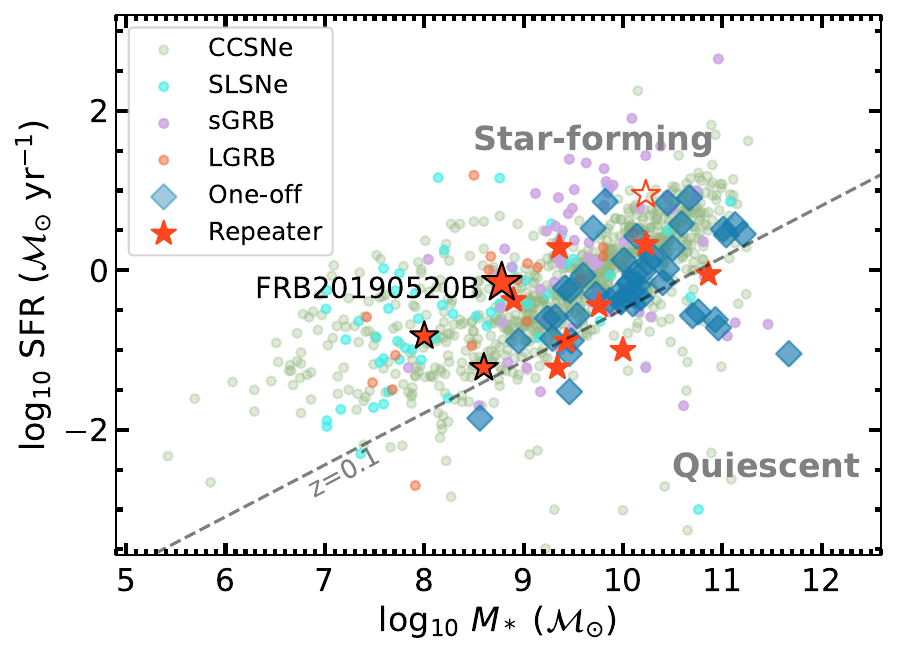} 
        \vspace{-0.2cm}
    \end{minipage}
    }

    \subfigure{
    \begin{minipage}[h]{\linewidth}
        \includegraphics[width=0.99\linewidth]{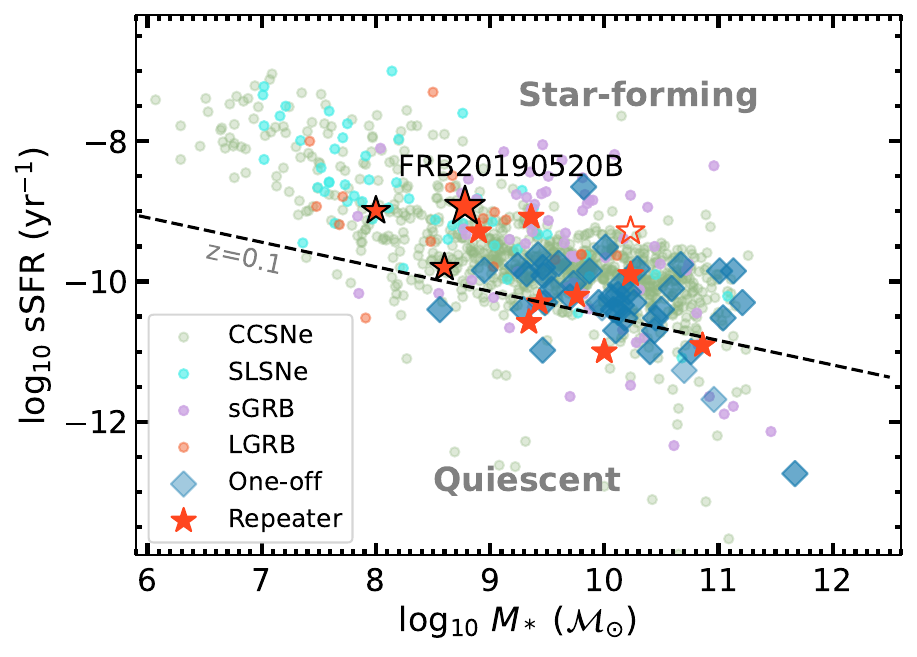}
        \vspace{-0.2cm}
    \end{minipage}
    }
    \caption{
    The relationship between stellar mass and SFR, as well as between stellar mass and sSFR, for the host galaxies of FRBs and other transients.
    Host galaxies of repeating FRBs are indicated by red stars, while one-off FRB host galaxies are shown as blue squares.
    The host galaxy of FRB 20190520B is represented by a larger star with a black edge.
    Small dots denote the host galaxies of CCSNe in gray, SLSNe in cyan, sGRBs in purple, and LGRBs in orange.
    The dashed line separates star-forming galaxies from quiescent galaxies at redshift $z=0.1$ \citep{Moustakas2013ApJ...767...50M}.
    Data references are provided in Section~\ref{subsect:SED}.
    }
    \label{fig:sfrM}
\end{figure}

Figure~\ref{fig:sfrM} illustrates the distribution of stellar mass and SFR, along with the stellar mass and sSFR, for the confirmed host galaxies of FRBs.
Additionally, for comparison, this figure also includes host galaxies of four other types of transients: core-collapse supernovae (CCSNe), superluminous supernovae (SLSNe), short gamma-ray bursts (sGRBs), and long gamma-ray bursts (LGRBs).
The CCSNe data is from \cite{CCSNe2021ApJS..255...29S}, the SLSNe and LGRBs data are provided by \cite{Taggart2021MNRAS.503.3931T}, and the sGRB data is provided by \cite{Nugent2022ApJ...940...57N}.

Comparing host galaxy properties of FRBs to those of other transients \citep[e.g.][]{Bhandari2020ApJ...895L..37B, Bhandari2101172022arXiv221116790B} and analyzing their global properties shed light on the progenitor population and formation mechanisms of FRBs. 
As discussed in Section \ref{sect:analy}, the host galaxy of FRB 20190520B is a low-mass, low-metallicity galaxy with a high SFR compared to the star formation main sequence at a similar redshift \citep{Moustakas2013ApJ...767...50M}.
In addition to being associated with a PRS, this galaxy shares many similarities with the host galaxy of FRB 20121102A. 
The high SFRs in both galaxies suggests the presence of young massive stars or young supernova remnants. 
Their low stellar mass, low metallicity, and high SFR place them in the same region as the host galaxies of LGRBs and SLSNe in Figure~\ref{fig:sfrM}, increasing the likelihood of extremely massive star progenitors, as also suggested by \cite{FengsigRM2022Sci...375.1266F}.

\section{Ionized Gas Properties} \label{sect:ionizedgas}

\subsection{Spatial Distribution} \label{subsect:gas_distrib}

\begin{figure*}[htb]
    \centering
    \subfigure[]{
    \begin{minipage}[h]{0.4\linewidth}
        \includegraphics[width=\textwidth]{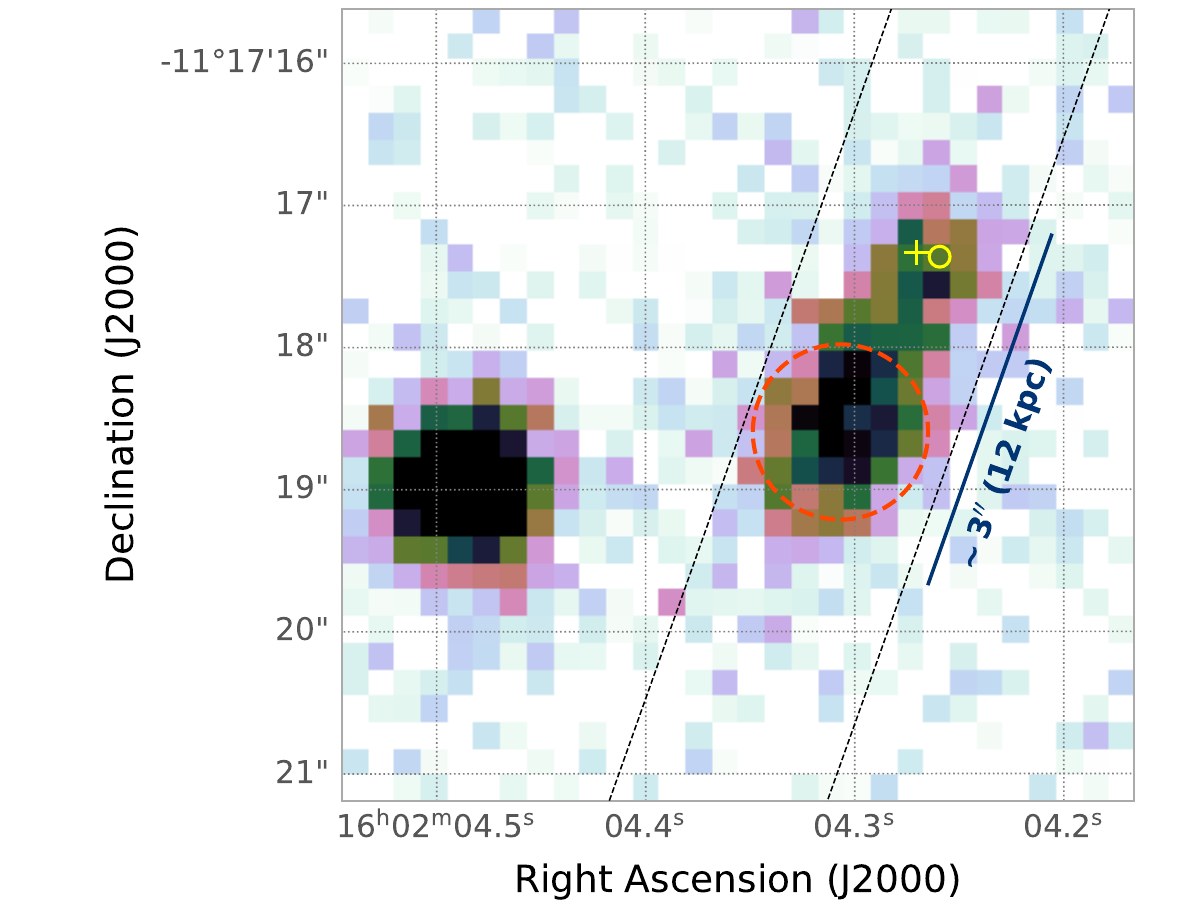}
    \end{minipage}
    }
    \subfigure[]{
    \begin{minipage}[h]{0.449\linewidth}
        \includegraphics[width=\textwidth]{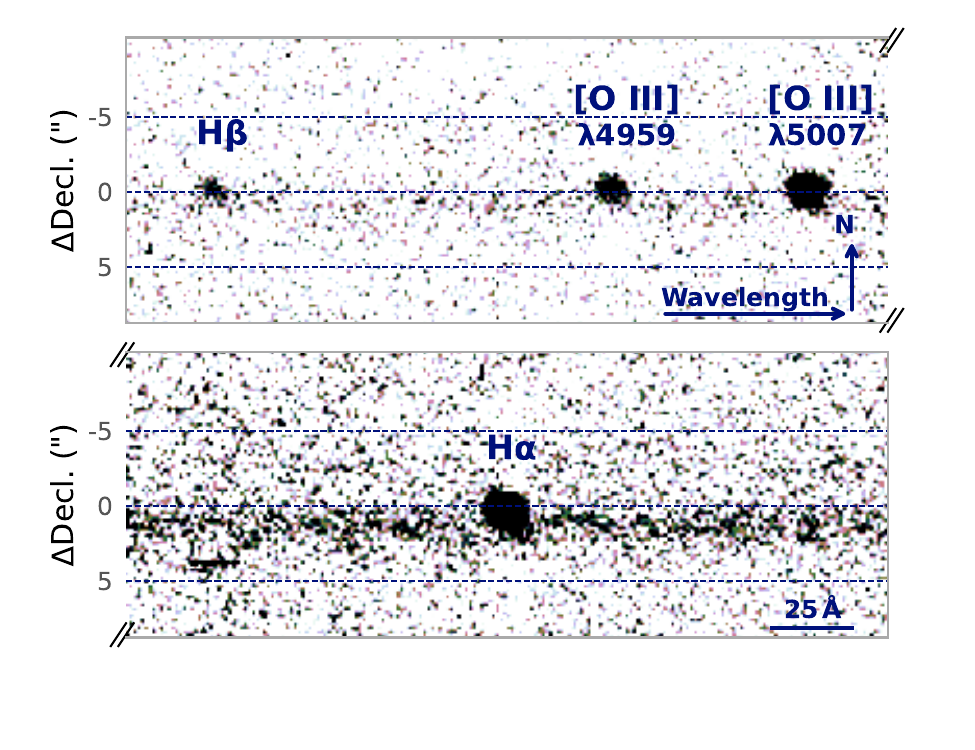}
    \end{minipage}
    }
    \caption{
    (a) CFHT \textit{R\arcm}-band image of the host galaxy of FRB 20190520B. The red dashed circle represent the PSF of the \textit{J}-band image. Two diagonal black dashed lines represent the Keck/LRIS long-slit, with a width of 1.5$^{\prime \prime}$. The yellow cross and circle indicate the locations of the FRB and PRS, respectively. The scale bar represents $3\arcsec$, or 12 kpc in projection at redshift $z=0.241$.
    (b) A portion of the 2-D spectrum obtained from the Keck/LRIS observations. The emission lines are labeled. ``N" indicates the northerly direction in the slit, which is oriented at 20 degrees from true north.
    ``Wavelength" indicates the wavelength, ranging from blue to red.
    $\Delta \text{Decl.}$ is the offset of the Gaussian fitting line center with respect to the center of the H$\alpha$ emission line.
    }
    \label{fig:2Dspec}
\end{figure*}

The \textit{J}-band image of the host galaxy of FRB 20190520B shows an unresolved near-infrared source \citep[panel b in Figure~2 of][]{Niu190520-2022Natur.606..873N}.
The source centroid is located approximately 1.4$^{\prime \prime}$ southeast of the FRB position, corresponding to a projected separation of 5.5 kpc at the host galaxy redshift.
The point-spread function (PSF) of the source is indicated by the red dashed circle in panel (a) of Figure~\ref{fig:2Dspec}. 
No strong emission lines fall within the bandpass of the \textit{J}-band for a galaxy at redshift $z = 0.241$, so the detected fluxes in the \textit{J}-band image are dominated by the stellar component of the host galaxy. 
The stellar continuum is clearly visible in the 2-D spectrum shown in Figure~\ref{fig:2Dspec}. 
Panel (b) of that figure reveals a distinct continuum component to the south of the center, while to the north, where the FRB and PRS are located, the dominance of ionized gas emission is observed.

\begin{figure}[htb]
    \centering
    \includegraphics[width=0.9\linewidth]{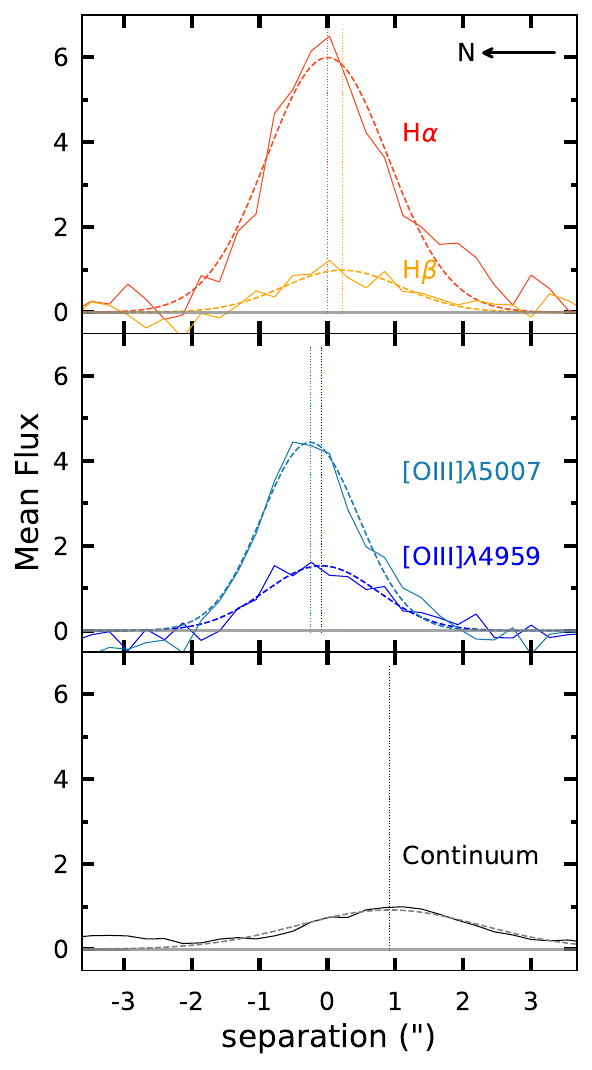}
    \caption{Spatial Gaussian fitting results of 2-D spectra of emission lines and continuum. The fitting results for the Balmer emission, [O III] emission and continuum are shown in the labeled top, medium, and bottom panels, respectively. The spatial direction is indicated by the arrow in the upper right corner of the top panel. The fitted results are detailed in Table~\ref{tab:line-gau}.
    }
    \label{fig:2D-gau}
\end{figure}

We fit the spatial distributions of the continuum and the emission lines separately along the slit using the entire red side 2-D spectrum, which ranges from 5440 to 10280 \AA. 
Before fitting the spatial distributions, we apply a correction to the trace of the 2-D spectrum using moment analysis to extract the spine of the trace, followed by fitting the trace with a 2nd-order polynomial. 
Figure~\ref{fig:2D-gau} shows the results of the Gaussian fitting, while Table~\ref{tab:line-gau} provides the best-fitting parameters.
Notably, the stellar continuum primarily extends toward the south, while the northernmost component is attributed to background contamination. 
The Balmer and [O III] emission are predominantly from the northern part of the galaxy. 
The slight offset observed between the best-fitting centers of the emission lines results from distortion in the 2D spectrum.
We find that the peaks of the stellar continuum and \Ha\ line emission are separated by $\sim 0.9^{\prime\prime}$, or 3.9 kpc in projection at position angle PA$=24.63^{\circ}$.
Furthermore, by separately fitting the emission lines located exclusively in the northern region and those overlapping with the continuum in the 2D spectrum, we found that the \Ha-derived radial velocity of the northern ionized gas is $(39.6 \pm 0.4)$ km s$^{-1}$ lower than that of the southern region.

\begin{table}[ht!]
    \centering
    \caption{Gaussian fitting parameters for the spatial distribution of the emission lines and continuum in the Keck/LRIS 2-D spectrum.}
    \medskip
    \begin{threeparttable}
    \begin{tabular}{lcc}
    \hline
    \hline
    \noalign{\smallskip}
    Component              & $\Delta$Decl.$^{\dagger}$ & FWHM \\
                           & (arcsec) & (arcsec) \\
    \noalign{\smallskip}
    \hline
    \noalign{\smallskip}
    H$\alpha$              & 0.00 & 2.11 \\
    \noalign{\smallskip}
    H$\beta$               & 0.22 & 2.04 \\
    \noalign{\smallskip}
    [O III] $\lambda$ 4959 & -0.09 & 1.95 \\
    \noalign{\smallskip}
    [O III] $\lambda$ 5007 & -0.26 & 1.68 \\
    \noalign{\smallskip}
    Continuum              & 0.91   & 3.20 \\
    \noalign{\smallskip}
    \hline
    \end{tabular}
    \medskip
    \begin{tablenotes}
    \footnotesize
    \item[$\dagger$] The Gaussian fitting line center is offset with respect to the center of the H$\alpha$ emission line.
    \end{tablenotes}
    \end{threeparttable}
    \label{tab:line-gau}
\end{table}

\begin{figure*}[htb]
    \centering
    \includegraphics[width=\linewidth]{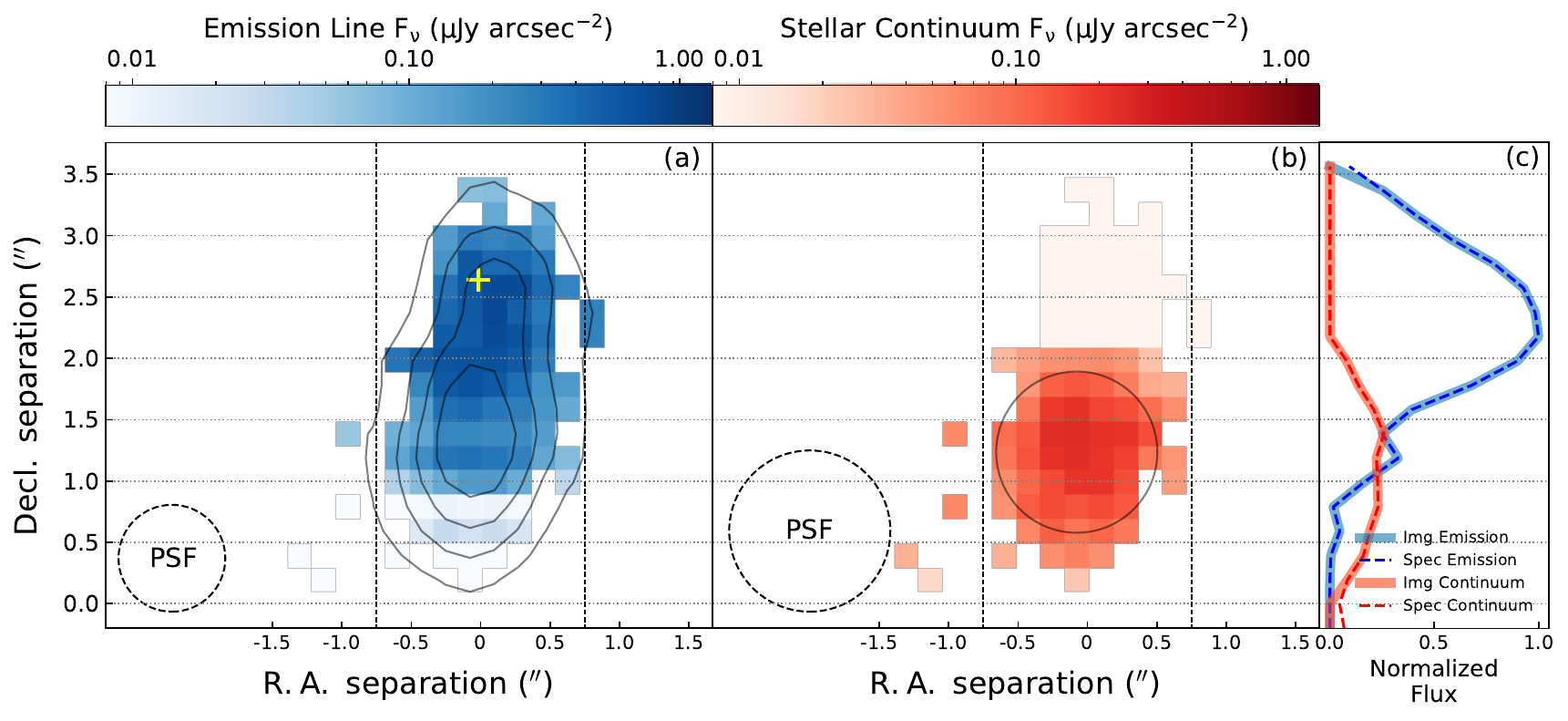}
    \caption{
    Emission region distribution of the FRB 20190520B host galaxy. 
    The left panel displays the distribution of the emission region. 
    The \textit{R\arcm}-band wavelength encompasses the H${\beta}$ and [O III]$\lambda4959/5007$ emission lines.
    The yellow crosses indicate the location of FRB 20190520B.
    The contours represent the \textit{R\arcm}-band image of the host galaxy.
    The middle panel shows the continuum residual in the \textit{R\arcm}-band. 
    The black circle over the image represents the unresolved region in the \textit{J}-band.
    The PSF size of the \textit{R\arcm}- and the \textit{J}-band are shown by the dashed circles with labels.
    The right panel presents the spatial mean flux profile of the emission and continuum components, respectively.}
    \label{fig:emissions}
\end{figure*}

The CFHT \textit{R\arcm}-band wavelength encompasses the H${\beta}$ and [O III]$\lambda4959/5007$ emission lines. 
To conduct a more detailed analysis of the distribution of ionized gas, we performed a decomposition of the emission component from the continuum within the wavelength range of 6000 to 6350 \AA, utilizing the results obtained from the 2-D spectral fitting. 
Figure~\ref{fig:emissions} illustrates the emission distribution of H${\beta}$ and [O III]$\lambda4959/5007$. 
The yellow cross indicates the location of the FRB, which aligns with the emission region primarily attributed to star-forming activities, as evidenced by the BPT diagram (Figure~\ref{fig:bpt}). 
The observed correlation between the location of FRB 20190520B and the distribution of ionized gas associated with star-forming regions suggests that this FRB may originate from young stellar populations, a notion further supported by \cite{FengsigRM2022Sci...375.1266F}.

We estimate the seeing-limited sSFR of the ionized region in the north. 
The stellar mass of the northern region within a diameter of 1.1$^{\prime \prime}$ is estimated using the Subaru \textit{J}-band image, yielding an upper limit of $(1.8 \pm 0.2) \times 10^8 \ M_{\odot}$. 
The flux of the northern H$\alpha$ emission line is estimated using the Keck 2D spectrum. 
By applying the ratio between two single-Gaussian profiles fitted for the northern and the entire H$\alpha$ emission, we estimate the lower limit of the H$\alpha$ emission flux to be $(5.9 \pm 0.1) \times 10^{-16}$ erg cm$^{-2}$ s$^{-1}$. 
Consequently, the H$\alpha$-traced star formation rate and specific star formation rate are estimated as SFR(H$\alpha$) = $0.6 \pm 0.1 \ M_{\odot} \ \rm yr^{-1}$ and $\rm log(sSFR/yr^{-1})(H\alpha) = -8.5 \pm 0.1 $, respectively. 
The sSFR of the northern part, which is dominated by ionized gas, is more than three times higher than that of the entire host galaxy. 
The overlap of the active FRB 20190520B with an active star-forming region in its host galaxy supports the scenario of a young stellar population as its progenitor.

\subsection{Contribution to DM} \label{subsect:halphaDM}

We utilize the H$\alpha$ emission line and its emission measurement (EM) to analyze the DM.
Assuming the size of the emission region to be $r=0.45^{\prime\prime}$, based on the unresolved CFHT/MegaCam image, the extinction-corrected H$\alpha$ surface density is $S_{\rm H\alpha} \approx 460 \pm 7$ Rayleighs (or $2.61 \pm 0.04$ erg cm$^{-2}$ s$^{-1}$ arcsec$^{-2}$) in the source frame at redshift $z=0.241$. 
This indicates that the EM in the source frame is

\begin{equation}
    \begin{aligned}
        \rm{EM_{\mathrm{H\alpha,s}}} &= 2.75 \, \mathrm{pc \ cm}^{-6} \, T_{4}^{0.9} \, S({\mathrm{H}\alpha})
        \\
        &\approx 1270 \pm 20 \ {\rm pc \ cm^{-6}} \cdot T_{4}^{0.9},
    \end{aligned}
\end{equation}

\noindent where $T_4$ is the temperature measured in units of $10^4$ K \citep{1977ApJ...216..433R}. 
The contribution of H$\alpha$ from the ionized gas in H$\alpha$ emission regions to the DM budget can be expressed as

\begin{equation}
    \begin{aligned}
            \rm{DM_{H\alpha}} &= \rm{EM_{H\alpha,s}^{1/2}} \cdot \it{l}^{\rm 1/2} \cdot \left [ \frac{\zeta(\rm 1+\epsilon^2)}{f} \right ]^{\rm -1/2}
            \\
            &\approx 1270 \pm 20 \ {\rm pc \ cm^{-6}} T_{4}^{0.9}  l^{1/2} \left [ \frac{\zeta(1+\epsilon^2)}{f} \right ]^{-1/2},
    \end{aligned}
\end{equation}

\noindent where $l$ denotes the path length through the gas sampled by the FRB \citep{2016arXiv160505890C}.
We adopt $l= 5$ kpc, consistent with \cite{Niu190520-2022Natur.606..873N}. 
The parameters $\zeta$, $\epsilon^2$, and $f$ are model parameters of the ionized cloudlet that represent the cloud-cloud variations in mean density ($\zeta \geq 1$), the variance of density fluctuations within a cloud ($0 \leq \epsilon^2 \leq 1$), and the filling factor ($0 \leq f \leq 1$), respectively. 
More than 90\% of the ISM exists in the diffuse ionized gas (DIG), or warm ionized medium (WIM) in the case of the Milky Way, which envelops the galaxy and is predominantly traced by \Ha\ emission \citep{WIM2009RvMP...81..969H}.
For the estimation, we adopt typical parameters of the Milky Way's WIM: $f = 0.1$, $\zeta = 1$, and $\epsilon^2 = 1$, with the [O III] emission yielding an electron temperature of $T_e = (3.2 \pm 1.6) \times 10^4$ K.
The estimated H$\alpha$-traced DM$_{\rm host}$ in the source frame is given by DM$_{H\alpha}^{s} \approx 950 \pm 220 \ {\rm pc \ cm^{-3}}$.
The observed DM$_{\rm host}$ is calculated as DM$_{\rm host} = {\rm DM_{H\alpha}^{s}} / (1 + z) = 760 \pm 180 \ {\rm pc \ cm^{-3}}$.
In the range of $1 \leq \zeta(1 + \epsilon^2)/f \leq 50$, the source frame DM$_{\rm host}$ could reach a maximum of $4300 \pm 1000 \ {\rm pc \ cm^{-3}}$ or a minimum of $600 \pm 150 \ {\rm pc \ cm^{-3}}$.

\cite{Niu190520-2022Natur.606..873N} analyzed the DM budget of FRB 20190520B and derived the observed DM$_{\rm host}$ to be between $\sim 745$ and $1020 \ \mathrm{pc \ cm^{-3}}$, which aligns with our findings. 
We suggest that the ionized ISM within the host galaxy contributes significantly to the high DM$_{\rm host}$ of FRB 20190520B.
This scenario is also supported by \cite{OckerHost2022ApJ...931...87O}, who propose a higher electron temperature than the typical value of $10^4$ K for warm ionized gas.
However, the effective path length of the FRB through the ionized gas remains uncertain, so the DM contributed by the local environment of the FRB cannot be excluded.

Although our analysis suggests that the ionized ISM of the host galaxy can account for the high \dmhost\ required, we also note that \cite{foreground2023ApJ...954L...7L} identified foreground galaxy clusters associated with the host galaxy of FRB 20190520B.
They reported that these two foreground galaxy clusters could contribute a combined $\rm DM_{halos} \sim 450 - 640 \ pc \ cm^{-3}$.

\section{Discussion} \label{sect:discussion}

\subsection{DM$_{\rm{host}}$ and DM$_{\rm{H\alpha}}$} \label{subsect:DMhost}

Analysis of DM$_{\rm{host}}$ reveals contributions from ionized gas in the host galaxy and further constrains the in-situ environment of the FRB. 
DM$_{\rm{host}}$ can be derived through the decomposition of the DM budget according to

\begin{equation}
\label{eq:DMs}
    \rm DM_{FRB} = DM_{MW} + DM_{cosmic} + DM_{host} + (DM_{foreGC}),
\end{equation}

\noindent where DM$_{\rm{FRB}}$ represents the total observed DM.
DM$_{\rm{MW}}$ arises from electrons in both the ISM of the Milky Way disk and halo.
The Galactic disk contribution can be estimated by the NE2001 model \citep{NE2001-12002astro.ph..7156C, NE2001-22003astro.ph..1598C} or the YMW16 model \citep{YMW2017ApJ...835...29Y}, while the Milky Way halo contribution is derived using \cite{DMhalo2020ApJ...888..105Y}.
DM$_{\rm{cosmic}}$ accounts for the DM originating from the intergalactic medium (IGM) along the line of sight. 
In this work, we estimate DM$_{\rm{cosmic}}$ approximately using the relation from \cite{ZhangDMz2018ApJ...867L..21Z} by:
\begin{equation}
    z \sim \rm DM_{cosmic}/855 pc \ cm^{-3},
\end{equation}
\noindent where $z$ is the redshift of the FRB.
DM$_{\rm{foreGC}}$ represents the DM contributions of foreground galaxy clusters, which remain under investigation for most FRBs.
Therefore, we do not include DM$_{\rm{foreGC}}$ in the current comparisons.
\dmhost\ includes contributions from the ionized gas in the host galaxy and the FRB's local environment.
In the following discussion, we focus on the DM contribution from the host galaxy, \dmhost, based on the analysis of the host galaxy's H$\alpha$ emission.
We denote DM$_{\rm host;H\alpha}$ as the DM calculated using the H$\alpha$ emission line from the host galaxy spectrum, as discussed in Section \ref{subsect:halphaDM}.

\begin{figure}[ht]
    \centering
    \includegraphics[width=0.99\linewidth]{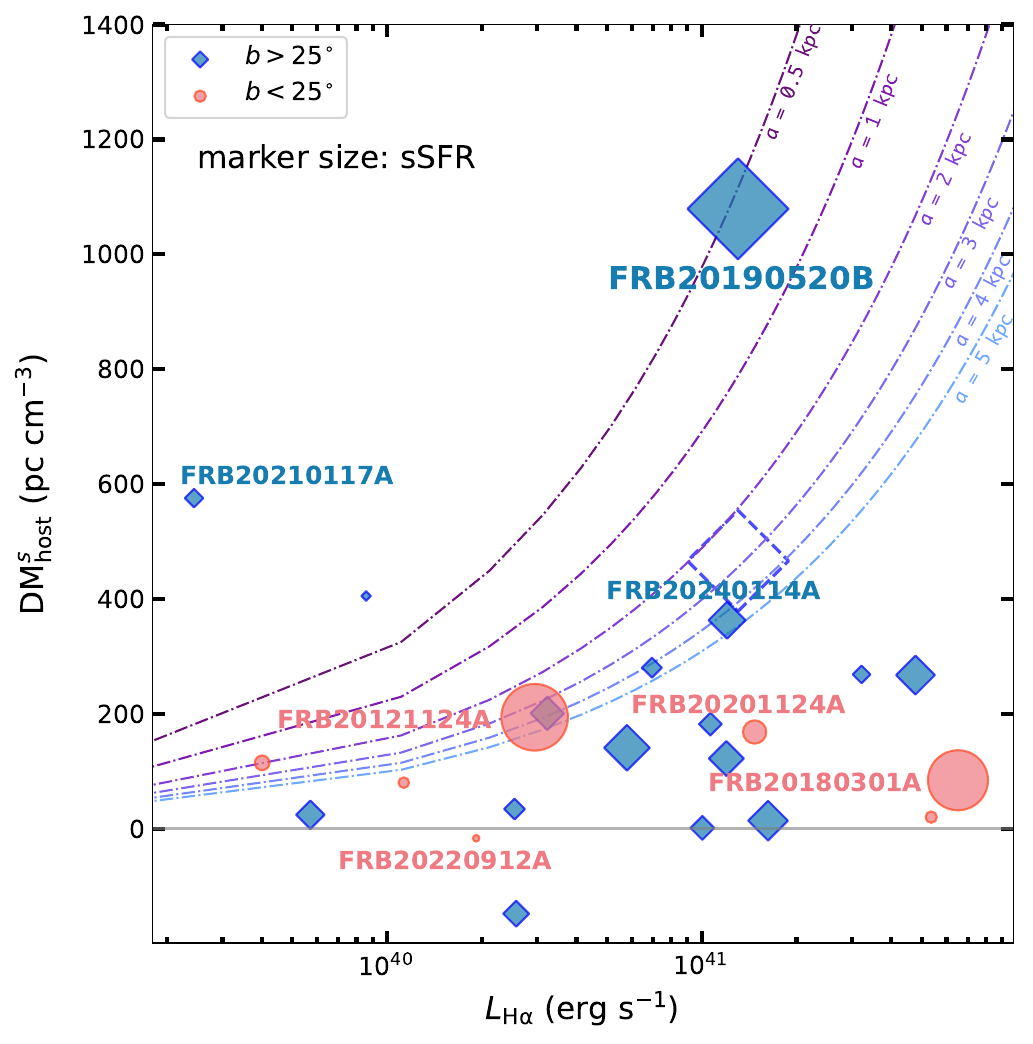}
    \caption{
    The relationship between H$\alpha$ luminosity of the host galaxy and the source-frame \dmhost. 
    The \dmhost \space are corrected for the Milky Way contribution according to \cite{NE2001-12002astro.ph..7156C, NE2001-22003astro.ph..1598C, DMhalo2020ApJ...888..105Y} and the IGM contribution using the DM-z scaling relation in \cite{ZhangDMz2018ApJ...867L..21Z}.
    The blue squares indicate FRBs at higher Galactic latitudes ($\lvert \rm b \rvert > 25^{\circ}$) and the red circles show sources at $\lvert \rm b \rvert<25^{\circ}$.
    Negative values in the plot may arise from insufficient modeling of DM$_{\rm MW}$ along the line of sight.
    The dash-dotted lines represent the equal-DM$_{\rm H\alpha}$ lines.
    The dashed square represents the DM$_{\rm host}^{s}$ of FRB 20190520B when the DM contributions of the foreground galaxy clusters are subtracted; see details in Sect.~\ref{subsect:DMhost}.
    }
    \label{fig:DM-Ha}
\end{figure}

We analyze the relationship between the ionized ISM of the host galaxy and \dmhost \space derived using Equation~\ref{eq:DMs}.
The NE2001 model is utilized to derive DM$_{\rm MW}$ for each source,
and all the derived data are listed in Table~\ref{tab:dm-ha-data}.
Figure~\ref{fig:DM-Ha} presents a scatter plot illustrating the correlation between DM$_{\rm{host}}$ (source frame) obtained from the DM budget and the H$\alpha$ luminosity $L_{\rm H\alpha}$ of the host galaxy spectra. 
For comparison, we draw the equal-DM$_{\rm H\alpha}$ lines assuming ionized gas cubes with side lengths and line-of-sight path lengths of $a = \textit{l} = 0.5, \ 1, \ 2, \ 3, \ 4, \ 5$ kpc, respectively, and a typical electron temperature of $T_e = 10,000$ K.
For FRB 20190520B, Figure~\ref{fig:DM-Ha} suggests that the high \dmhost\ could be attributed to a small-scale ionized region with $T_e = 10,000$ K. Alternatively, as discussed in Sect.~\ref{subsect:halphaDM} and \cite{OckerHost2022ApJ...931...87O}, a higher electron density could also contribute significantly to \dmhost.
We also note that \cite{OckerScatter2023MNRAS.519..821O} attributed the variation in scattering time observed in FRB 20190520B to a dynamic and inhomogeneous plasma environment surrounding the FRB.
This local environment could also contribute to the substantial excess in \dmhost.
However, when the DM contributions from foreground galaxy clusters are considered, the source frame DM$_{\rm{host}}$ could be reduced to $466.5^{+139.7}_{-230.1}$ pc cm$^{-3}$ \citep[blue dashed square in Figure~\ref{fig:DM-Ha}, ][]{foreground2023ApJ...954L...7L}.
This scenario is consistent with a longer effective path length of a few kpc with a lower $T_e$.
Given the large uncertainties of the $T_e$ measurements and the off-center distribution of the ionized gas within the host galaxy, we cannot rule out the possibility that the DM$_{\rm{host}}$ is primarily contributed by the ionized gas of the host galaxy on kpc scales.

For other FRBs, \cite{Tendulkar121102-2017ApJ...834L...7T} found that for FRB 20121102A, the budget-derived DM$_{\rm{host}} < $DM$_{\rm host;H\alpha}$ which is estimated using the extinction corrected \Ha\ emission, suggesting that the ionized region in the host galaxy exhibits some clumpiness or that the effective path length is smaller than the size of the ionized region. 
In contrast, for FRB 20210117A, DM$_{\rm host;H\alpha}$ is significantly lower than budget-derived DM$_{\rm{host}}$, indicating excess DM from the FRB's local environment \citep{Bhandari2101172022arXiv221116790B}. 

For half of our samples, the budget-derived \dmhost\ is lower than DM$_{\rm H\alpha}$. This discrepancy could result from a less dense or clumpy ionized ISM in the FRB's local environment, a shorter effective path along the FRB sightline, or the distinct location of the FRBs relative to the ionized gas regions in their host galaxies.

\subsection{PRS} \label{subsect:prs}

Upon the discovery of FRB 20190520B and the associated PRS, their estimated locations had a projected separation of 0.165$^{\prime \prime}$, equivalent to 0.65 kpc in projection at $z=0.241$.
\cite{Niu190520-2022Natur.606..873N} concluded that they are physically connected. 
Unresolved VLA observations indicated that the PRS has a size smaller than 1.4 kpc. 
In a recent study, \cite{190520PRS_VLBI} conducted very-long-baseline interferometry (VLBI) observations of the compact PRS associated with FRB 20190520B, constraining its size to be smaller than 9 pc.
During these PRS observations, a simultaneous detection of an FRB burst occurred, demonstrating that the FRB's position consistently aligns with the PRS location within $\leq 20$ mas, corresponding to $\leq 80$ pc in the source frame.
This detection strongly reinforces the physical association between the PRS and the FRB, supporting the hypothesis that a single central engine powers both the bursts and the PRS.
The origin of the PRS could potentially be a pulsar wind nebula \citep{YD2019ApJ...885..149Y}, a synchrotron nebula heated by FRB emission \citep{LYD2020ApJ...896...71L}, or an accreting massive black hole with low luminosity \citep{Michilli201102-2018Natur.553..182M}.
Additionally, \cite{Eftekhari2019} detected an unresolved radio source correlated with the Type I SLSNe PTF10hgi 7.5 years after its explosion, implying that the PRS could be due to delayed radio emission from a SLSNe.

\cite{2024ApJ...976..165Y} analyzed the flux and variability of the PRS associated with FRB 20190520B.
The SFR derived from radio emission, estimated using the flux density of the PRS, is several dozen times higher than the SFR derived from H$\alpha$, potentially indicating the presence of a highly obscured star-forming region.
However, in our analysis, the electron temperature ($T_e$) of the ionized gas in the host galaxy is higher than that of typical warm plasma, and the estimated $T_e$ represents a lower limit.
Both the elevated $T_e$ and the low metallicity of the host galaxy are inconsistent with a highly dust-obscured scenario.
This implies that the radio-traced highly obscured star-forming region could either be compact source embedded within the ionized region, or that the radio emission is not primarily due to star formation.

\section{Summary} \label{sect:sum}

In this study, we conducted a comprehensive analysis of the host galaxy of FRB 20190520B, focusing on its ionized gas properties, SFR, and the implications for the observed DM.
The host galaxy is a low-metallicity galaxy, with an oxygen abundance of $12+ {\rm log_{10} \ ([O/H])} \geq 7.4 \pm 0.1 $, exhibiting strong emission features at a redshift of $z=0.241 \pm 0.001$.
The intrinsic extinction is measured to be $A_V = 1.24 \pm 0.31$.
The stellar mass, derived from \textit{J}-band photometry, is calculated to be \((6.2 \pm 0.8) \times 10^8 \, M_{\odot}\).
The SFR derived from the H$\alpha$ luminosity is $\sim 0.7 \ M_{\odot} \ \rm{yr^{-1}}$, while the specific star formation rate is determined to be $\rm{log \ sSFR/yr^{-1}} \sim -9.0 $.

Using two-dimensional spectral fitting from Keck/LRIS, we mapped the spatial distribution of the continuum and emission lines.
This reveals that significant Balmer and [O III] emission are concentrated in the northern region associated with the FRB and the PRS, indicating active star formation, while the stellar continuum predominantly extends to the southern part of the host galaxy.
The ionized region exhibits a high sSFR of $\rm log(sSFR/yr^{-1})(H\alpha) = -8.5 \pm 0.1 $, which is more than three times higher than that of the entire galaxy.
This elevated sSFR underscores the presence of an active star-forming environment in the northern part of the host galaxy.

We further investigated the contribution of the ionized ISM to the DM of FRB 20190520B.
Our analysis indicates that ionized gas plays a significant role in contributing to the high observed DM.
The derived electron temperature $T_e$ from optical emission lines, including [O III]$\lambda4363$, suggests a warm ionized medium with a higher $T_e$ than typical values for warm plasma.
The estimated DM from H$\alpha$ emission in the source frame is $950 \pm 220 \ \text{pc cm}^{-3}$, consistent with previous studies.

Furthermore, we investigated whether the PRS linked to the FRB indicates the presence of a highly obscured star forming region.
The high electron temperature and low metallicity of the host galaxy challenge the likelihood of such a scenario.
The unresolved scale of the PRS, approximately 9 pc, leaves questions about whether the radio emission primarily originates from a extremely active and compact star forming region or other processes.

\vspace{1cm}
C.-W.T., X.-L.C. and M.L. acknowledge support from NSFC No. 11988101, No. 12041302, and the International Partnership Program of Chinese Academy of Sciences, Program No.114A11KYSB20210010. 
The authors would like to thank the referee for their useful comments on the manuscript.
This work is supported by CAS project No. JZHKYPT-2021-06.
The work of DS was carried out at the Jet Propulsion Laboratory, California Institute of Technology, under a contract with the National Aeronautics and Space Administration (80NM0018D0004).
Di Li is a New Cornerstone Investigator.
C.-H.N. is supported by NSFC No.12203069 and the CAS Youth Interdisciplinary Team and the Foundation of Guizhou Provincial Education Department for Grants No. KY(2023)059. R.J.A. was supported by FONDECYT grant number 1231718 and by the ANID BASAL project FB210003.

This work is based
on observations obtained with MegaPrime/MegaCam, a joint project of CFHT and CEA/DAPNIA, at the Canada-France-Hawaii Telescope (CFHT) which is operated by the National Research Council (NRC) of Canada, the Institut National des Science de l'Univers of the Centre National de la Recherche Scientifique (CNRS) of France, and the University of Hawaii. The observations at the Canada-France-Hawaii Telescope were performed with care and respect from the summit of Maunakea which is a significant cultural and historic site. 

This research is based in part on data collected at the Subaru Telescope, which is operated by the National Astronomical Observatory of Japan. We are honored and grateful for the opportunity of observing the Universe from Maunakea, which has the cultural, historical, and natural significance in Hawaii. 

Some of the data presented herein were obtained at the W. M. Keck Observatory, which is operated as a scientific partnership among the California Institute of Technology, the University of California and the National Aeronautics and Space Administration. The Observatory was made possible by the generous financial support of the W. M. Keck Foundation. 

Based on observations made with the Gran Telescopio Canarias (GTC), installed at the Spanish Observatorio del Roque de los Muchachos of the Instituto de Astrofísica de Canarias, on the island of La Palma.

The authors wish to recognize and acknowledge the very significant cultural role and reverence that the summit of Maunakea has always had within the indigenous Hawaiian community.  We are most fortunate to have the opportunity to conduct observations from this mountain.

\vspace{5mm}
\facilities{CFHT(MagaCam), Subaru(MOIRCS), Hale(DBSP), Keck:I(LRIS), GTC(OSIRIS$+$)}

\software{
astropy \citep{Astropy2022ApJ...935..167A},
ccdproc \citep{ccdprocmatt_craig_2017_1069648},
emcmc \citep{emcmc2013PASP..125..306F},
photutils \citep{larry_bradley_2024_12585239},
scipy \citep{2020SciPy-NMeth},
DrizzlePac \citep{drizzle2021AAS...23821602H},
IRAF \citep{IRAF1986SPIE..627..733T, IRAF1993ASPC...52..173T}
}

\clearpage
\appendix
\renewcommand{\thetable}{\Alph{section}.\arabic{table}}
\setcounter{table}{0}
\vspace{-0.5cm}
\section{DM Budget and $L_{\rm{H\alpha}}$}
\vspace{-0.25cm}
\input{dm_ha_data.tex}

\bibliography{sample631}{}
\bibliographystyle{aasjournal}

\end{document}

%% file: dm_ha_data.tex
\begin{table}[htbp]
    \caption{Derived DM$_{\rm host}$ (source frame) from the DM budget and H$\alpha$ luminosity of the FRB host galaxy spectra, as shown in Figure~\ref{fig:DM-Ha} in Sect.~\ref{subsect:DMhost}. }
    \medskip
    \centering
    \begin{threeparttable}
    \begin{tabularx}{\linewidth}{lcccccccc}
    \hline
    \hline
    \noalign{\smallskip}
Name & $z_{\rm spec}$ & DM$_{\rm obs}$ & Ref.$^{\dagger}$ & DM$_{\rm MW\,disk;\,NE2001}$ & DM$_{\rm MW\,halo}$ & DM$_{\rm IGM}$ & DM$_{\rm host}^{s;\ast}$& $L_{\rm H\alpha}$ \\
 &  & (pc cm$^{-3}$) &  & (pc cm$^{-3}$) & (pc cm$^{-3}$) & (pc cm$^{-3}$) & (pc cm$^{-3}$) & ($10 ^{40}$ erg s$^{-1}$) \\
\hline
FRB20121102A & 0.1927 & 557 & (1,2) & 188.43 & 40.91 & 164.76 & 194.29 & 2.95 \\
FRB20180301A & 0.3305 & 536 & (3) & 151.72 & 38.06 & 282.58 & 84.68 & 65.22 \\
FRB20180916B & 0.0337 & 348.8 & (4) & 198.96 & 43.08 & 28.81 & 80.58 & 1.13 \\
FRB20180924B & 0.3214 & 362.16 & (5) & 40.5 & 45.54 & 274.8 & 1.75 & 10.05 \\
FRB20181112A & 0.4755 & 589 & (6) & 41.72 & 45.04 & 406.55 & 141.18 & 5.79 \\
FRB20190102C & 0.2913 & 364.54 & (7) & 57.4 & 46.99 & 249.06 & 14.33 & 16.25 \\
FRB20190520B & 0.241 & 1205 & (8) & 60.21 & 69.16 & 206.06 & 1079.14 & 13.05 \\
FRB20190608B & 0.1178 & 340.05 & (9,10) & 37.27 & 38.89 & 100.72 & 182.39 & 10.66 \\
FRB20190611B & 0.3778 & 321 & (4) & 57.83 & 47.21 & 323.02 & -147.51 & 2.57 \\
FRB20190714A & 0.2365 & 504.13 & (11) & 38.49 & 36.44 & 202.21 & 280.68 & 6.95 \\
FRB20191001A & 0.234 & 507.9 & (11) & 44.17 & 46.71 & 200.07 & 267.71 & 47.78 \\
FRB20191228A & 0.2432 & 297.5 & (11) & 32.95 & 36.88 & 207.94 & 24.54 & 0.57 \\
FRB20200430A & 0.1608 & 380.25 & (11) & 27.18 & 42.24 & 137.48 & 201.22 & 3.23 \\
FRB20200906A & 0.3688 & 577.8 & (11) & 35.84 & 30.16 & 315.32 & 268.94 & 32.2 \\
FRB20201124A & 0.0979 & 413.52 & (12) & 139.95 & 36.24 & 83.7 & 168.67 & 14.73 \\
FRB20210117A & 0.2145 & 728.95 & (11) & 34.38 & 37.17 & 183.4 & 575.69 & 0.24 \\
FRB20210410D & 0.1415 & 578.78 & (13) & 56.19 & 46.48 & 120.98 & 405.37 & 0.86 \\
FRB20220207C & 0.04304 & 262.38 & (14,15) & 76.1 & 39.21 & 36.8 & 115.02 & 0.4 \\
FRB20220208A & 0.351 & 440.73 & (15) & 101.56 & 41.46 & 300.1 & -3.24 & 239.6 \\
FRB20220307B & 0.248123 & 499.27 & (14,15) & 128.23 & 41.27 & 212.15 & 146.81 & 108.32 \\
FRB20220330D & 0.3714 & 467.79 & (15) & 38.54 & 31.22 & 317.55 & 110.37 & 37.73 \\
FRB20220506D & 0.30039 & 396.97 & (14,15) & 84.56 & 39.75 & 256.83 & 20.58 & 53.62 \\
FRB20220726A & 0.3619 & 686.23 & (15) & 89.52 & 35.32 & 309.42 & 343.16 & 8.12 \\
FRB20220912A & 0.0771 & 219.46 & (16) & 125.2 & 43.44 & 65.92 & -16.27 & 1.92 \\
FRB20220920A & 0.158239 & 314.99 & (14,15) & 39.86 & 34.12 & 135.29 & 122.45 & 11.98 \\
FRB20221101B & 0.2395 & 491.55 & (15) & 131.25 & 42.13 & 204.77 & 140.57 & 0.63 \\
FRB20221113A & 0.2505 & 411.03 & (15) & 91.73 & 35.54 & 214.18 & 87.01 & 131.44 \\
FRB20221116A & 0.2764 & 643.45 & (15) & 132.26 & 39.85 & 236.32 & 299.98 & 0.57 \\
FRB20230124A & 0.0939 & 590.57 & (15) & 38.61 & 33.58 & 80.28 & 479.23 & 103.06 \\
FRB20230307A & 0.2706 & 608.85 & (15) & 37.58 & 31.51 & 231.36 & 391.86 & 0.99 \\
FRB20230501A & 0.3015 & 532.47 & (15) & 125.7 & 41.9 & 257.78 & 139.37 & 43.2 \\
FRB20230626A & 0.327 & 452.72 & (15) & 39.25 & 33.81 & 279.59 & 132.8 & 51.37 \\
FRB20230628A & 0.127 & 344.95 & (15) & 39 & 31.44 & 108.58 & 187 & 91.41 \\
FRB20231120A & 0.0368 & 437.74 & (15) & 43.81 & 31.62 & 31.46 & 343.02 & 4.3 \\
FRB20231123B & 0.2621 & 396.86 & (15) & 40.31 & 34.28 & 224.1 & 123.9 & 2.89 \\
FRB20240114A & 0.1306 & 527.7 & (17,18) & 49.67 & 45.05 & 111.66 & 363.28 & 12.04 \\
\hline
    \end{tabularx}
    \medskip
    \begin{tablenotes}
    \footnotesize 
    \item[$\dagger$] DM$_{\rm obs}$ and redshift references of FRBs: (1) \cite{Chatterjee121102-2017Natur.541...58C} (2) \cite{Tendulkar121102-2017ApJ...834L...7T} (3) \cite{Luo1803012020Natur.586..693L} (4) \cite{1809162020Natur.577..190M} (5) \cite{1809242019Sci...365..565B} (6) \cite{1811122019Sci...366..231P}  (7) \cite{MacquartRelation2020Natur.581..391M} (8) \cite{Niu190520-2022Natur.606..873N} (9) \cite{1906082020MNRAS.497.3335D} (10) \cite{1906082020ApJ...895L..37B} (11) \cite{Bhandari2022AJ....163...69B} (12) \cite{201124Ravi2022MNRAS.513..982R} (13) \cite{Caleb2023MNRAS.524.2064C} (14) \cite{LawDSA2024ApJ...967...29L} (15) \cite{Sharma2024Natur.635...61S} (16) \cite{220912Ravi2023ApJ...949L...3R} (17) \cite{Tian2024MNRAS.533.3174T} (18) \cite{chen2025ApJ...980L..24C}
    \item [$\ast$] DM$_{\rm host}^{s} = $ DM$_{\rm host} \times (1+z)$, where DM$_{\rm host}$ is the DM budget-derived value and DM$_{\rm host}^{s}$ represents the source frame value.
    \end{tablenotes}
    \end{threeparttable}
    \label{tab:dm-ha-data}
\end{table}

%% file: main631-v202411_arxiv_v2.bbl
\begin{thebibliography}{}
\expandafter\ifx\csname natexlab\endcsname\relax\def\natexlab#1{#1}\fi
\providecommand{\url}[1]{\href{#1}{#1}}
\providecommand{\dodoi}[1]{doi:~\href{http://doi.org/#1}{\nolinkurl{#1}}}
\providecommand{\doeprint}[1]{\href{http://ascl.net/#1}{\nolinkurl{http://ascl.net/#1}}}
\providecommand{\doarXiv}[1]{\href{https://arxiv.org/abs/#1}{\nolinkurl{https://arxiv.org/abs/#1}}}

\bibitem[{{Abdurro'uf} {et~al.}(2022){Abdurro'uf}, {Accetta}, {Aerts}, {Silva Aguirre}, {Ahumada}, {Ajgaonkar}, {Filiz Ak}, {Alam}, {Allende Prieto}, {Almeida}, {Anders}, {Anderson}, {Andrews}, {Anguiano}, {Aquino-Ort{\'\i}z}, {Arag{\'o}n-Salamanca}, {Argudo-Fern{\'a}ndez}, {Ata}, {Aubert}, {Avila-Reese}, {Badenes}, {Barb{\'a}}, {Barger}, {Barrera-Ballesteros}, {Beaton}, {Beers}, {Belfiore}, {Bender}, {Bernardi}, {Bershady}, {Beutler}, {Bidin}, {Bird}, {Bizyaev}, {Blanc}, {Blanton}, {Boardman}, {Bolton}, {Boquien}, {Borissova}, {Bovy}, {Brandt}, {Brown}, {Brownstein}, {Brusa}, {Buchner}, {Bundy}, {Burchett}, {Bureau}, {Burgasser}, {Cabang}, {Campbell}, {Cappellari}, {Carlberg}, {Wanderley}, {Carrera}, {Cash}, {Chen}, {Chen}, {Cherinka}, {Chiappini}, {Choi}, {Chojnowski}, {Chung}, {Clerc}, {Cohen}, {Comerford}, {Comparat}, {da Costa}, {Covey}, {Crane}, {Cruz-Gonzalez}, {Culhane}, {Cunha}, {Dai}, {Damke}, {Darling}, {Davidson}, {Davies}, {Dawson}, {De Lee}, {Diamond-Stanic}, {Cano-D{\'\i}az}, {S{\'a}nchez},
  {Donor}, {Duckworth}, {Dwelly}, {Eisenstein}, {Elsworth}, {Emsellem}, {Eracleous}, {Escoffier}, {Fan}, {Farr}, {Feng}, {Fern{\'a}ndez-Trincado}, {Feuillet}, {Filipp}, {Fillingham}, {Frinchaboy}, {Fromenteau}, {Galbany}, {Garc{\'\i}a}, {Garc{\'\i}a-Hern{\'a}ndez}, {Ge}, {Geisler}, {Gelfand}, {G{\'e}ron}, {Gibson}, {Goddy}, {Godoy-Rivera}, {Grabowski}, {Green}, {Greener}, {Grier}, {Griffith}, {Guo}, {Guy}, {Hadjara}, {Harding}, {Hasselquist}, {Hayes}, {Hearty}, {Hern{\'a}ndez}, {Hill}, {Hogg}, {Holtzman}, {Horta}, {Hsieh}, {Hsu}, {Hsu}, {Huber}, {Huertas-Company}, {Hutchinson}, {Hwang}, {Ibarra-Medel}, {Chitham}, {Ilha}, {Imig}, {Jaekle}, {Jayasinghe}, {Ji}, {Johnson}, {Jones}, {J{\"o}nsson}, {Katkov}, {Khalatyan}, {Kinemuchi}, {Kisku}, {Knapen}, {Kneib}, {Kollmeier}, {Kong}, {Kounkel}, {Kreckel}, {Krishnarao}, {Lacerna}, {Lane}, {Langgin}, {Lavender}, {Law}, {Lazarz}, {Leung}, {Leung}, {Lewis}, {Li}, {Li}, {Lian}, {Liang}, {Lin}, {Lin}, {Lin}, {Lintott}, {Long}, {Longa-Pe{\~n}a}, {L{\'o}pez-Cob{\'a}}, {Lu},
  {Lundgren}, {Luo}, {Mackereth}, {de la Macorra}, {Mahadevan}, {Majewski}, {Manchado}, {Mandeville}, {Maraston}, {Margalef-Bentabol}, {Masseron}, {Masters}, {Mathur}, {McDermid}, {Mckay}, {Merloni}, {Merrifield}, {Meszaros}, {Miglio}, {Di Mille}, {Minniti}, {Minsley}, {Monachesi}, {Moon}, {Mosser}, {Mulchaey}, {Muna}, {Mu{\~n}oz}, {Myers}, {Myers}, {Nadathur}, {Nair}, {Nandra}, {Neumann}, {Newman}, {Nidever}, {Nikakhtar}, {Nitschelm}, {O'Connell}, {Garma-Oehmichen}, {Luan Souza de Oliveira}, {Olney}, {Oravetz}, {Ortigoza-Urdaneta}, {Osorio}, {Otter}, {Pace}, {Padilla}, {Pan}, {Pan}, {Parikh}, {Parker}, {Peirani}, {Pe{\~n}a Ram{\'\i}rez}, {Penny}, {Percival}, {Perez-Fournon}, {Pinsonneault}, {Poidevin}, {Poovelil}, {Price-Whelan}, {B{\'a}rbara de Andrade Queiroz}, {Raddick}, {Ray}, {Rembold}, {Riddle}, {Riffel}, {Riffel}, {Rix}, {Robin}, {Rodr{\'\i}guez-Puebla}, {Roman-Lopes}, {Rom{\'a}n-Z{\'u}{\~n}iga}, {Rose}, {Ross}, {Rossi}, {Rubin}, {Salvato}, {S{\'a}nchez}, {S{\'a}nchez-Gallego}, {Sanderson}, {Santana
  Rojas}, {Sarceno}, {Sarmiento}, {Sayres}, {Sazonova}, {Schaefer}, {Schiavon}, {Schlegel}, {Schneider}, {Schultheis}, {Schwope}, {Serenelli}, {Serna}, {Shao}, {Shapiro}, {Sharma}, {Shen}, {Shetrone}, {Shu}, {Simon}, {Skrutskie}, {Smethurst}, {Smith}, {Sobeck}, {Spoo}, {Sprague}, {Stark}, {Stassun}, {Steinmetz}, {Stello}, {Stone-Martinez}, {Storchi-Bergmann}, {Stringfellow}, {Stutz}, {Su}, {Taghizadeh-Popp}, {Talbot}, {Tayar}, {Telles}, {Teske}, {Thakar}, {Theissen}, {Tkachenko}, {Thomas}, {Tojeiro}, {Hernandez Toledo}, {Troup}, {Trump}, {Trussler}, {Turner}, {Tuttle}, {Unda-Sanzana}, {V{\'a}zquez-Mata}, {Valentini}, {Valenzuela}, {Vargas-Gonz{\'a}lez}, {Vargas-Maga{\~n}a}, {Alfaro}, {Villanova}, {Vincenzo}, {Wake}, {Warfield}, {Washington}, {Weaver}, {Weijmans}, {Weinberg}, {Weiss}, {Westfall}, {Wild}, {Wilde}, {Wilson}, {Wilson}, {Wilson}, {Wolf}, {Wood-Vasey}, {Yan}, {Zamora}, {Zasowski}, {Zhang}, {Zhao}, {Zheng}, {Zheng}, \& {Zhu}}]{SDSSDR172022ApJS..259...35A}
{Abdurro'uf}, {Accetta}, K., {Aerts}, C., {et~al.} 2022, \apjs, 259, 35, \dodoi{10.3847/1538-4365/ac4414}

\bibitem[{{Aller}(1984)}]{Aller1984ASSL..112.....A}
{Aller}, L.~H. 1984, {Physics of thermal gaseous nebulae} (Springer Dordrecht), \dodoi{10.1007/978-94-010-9639-3}

\bibitem[{{Anna-Thomas} {et~al.}(2023){Anna-Thomas}, {Connor}, {Dai}, {Feng}, {Burke-Spolaor}, {Beniamini}, {Yang}, {Zhang}, {Aggarwal}, {Law}, {Li}, {Niu}, {Chatterjee}, {Cruces}, {Duan}, {Filipovic}, {Hobbs}, {Lynch}, {Miao}, {Niu}, {Ocker}, {Tsai}, {Wang}, {Xue}, {Yao}, {Yu}, {Zhang}, {Zhang}, {Zhu}, \& {Zhu}}]{reshma23}
{Anna-Thomas}, R., {Connor}, L., {Dai}, S., {et~al.} 2023, Science, 380, 599, \dodoi{10.1126/science.abo6526}

\bibitem[{{Asplund} {et~al.}(2021){Asplund}, {Amarsi}, \& {Grevesse}}]{solar3-2021A&A...653A.141A}
{Asplund}, M., {Amarsi}, A.~M., \& {Grevesse}, N. 2021, \aap, 653, A141, \dodoi{10.1051/0004-6361/202140445}

\bibitem[{{Assef} {et~al.}(2010){Assef}, {Kochanek}, {Brodwin}, {Cool}, {Forman}, {Gonzalez}, {Hickox}, {Jones}, {Le Floc'h}, {Moustakas}, {Murray}, \& {Stern}}]{2010ApJ...713..970A}
{Assef}, R.~J., {Kochanek}, C.~S., {Brodwin}, M., {et~al.} 2010, \apj, 713, 970, \dodoi{10.1088/0004-637X/713/2/970}

\bibitem[{{Assef} {et~al.}(2015){Assef}, {Eisenhardt}, {Stern}, {Tsai}, {Wu}, {Wylezalek}, {Blain}, {Bridge}, {Donoso}, {Gonzales}, {Griffith}, \& {Jarrett}}]{2015ApJ...804...27A}
{Assef}, R.~J., {Eisenhardt}, P.~R.~M., {Stern}, D., {et~al.} 2015, \apj, 804, 27, \dodoi{10.1088/0004-637X/804/1/27}

\bibitem[{{Astropy Collaboration} {et~al.}(2022){Astropy Collaboration}, {Price-Whelan}, {Lim}, {Earl}, {Starkman}, {Bradley}, {Shupe}, {Patil}, {Corrales}, {Brasseur}, {N{\"o}the}, {Donath}, {Tollerud}, {Morris}, {Ginsburg}, {Vaher}, {Weaver}, {Tocknell}, {Jamieson}, {van Kerkwijk}, {Robitaille}, {Merry}, {Bachetti}, {G{\"u}nther}, {Aldcroft}, {Alvarado-Montes}, {Archibald}, {B{\'o}di}, {Bapat}, {Barentsen}, {Baz{\'a}n}, {Biswas}, {Boquien}, {Burke}, {Cara}, {Cara}, {Conroy}, {Conseil}, {Craig}, {Cross}, {Cruz}, {D'Eugenio}, {Dencheva}, {Devillepoix}, {Dietrich}, {Eigenbrot}, {Erben}, {Ferreira}, {Foreman-Mackey}, {Fox}, {Freij}, {Garg}, {Geda}, {Glattly}, {Gondhalekar}, {Gordon}, {Grant}, {Greenfield}, {Groener}, {Guest}, {Gurovich}, {Handberg}, {Hart}, {Hatfield-Dodds}, {Homeier}, {Hosseinzadeh}, {Jenness}, {Jones}, {Joseph}, {Kalmbach}, {Karamehmetoglu}, {Ka{\l}uszy{\'n}ski}, {Kelley}, {Kern}, {Kerzendorf}, {Koch}, {Kulumani}, {Lee}, {Ly}, {Ma}, {MacBride}, {Maljaars}, {Muna}, {Murphy}, {Norman},
  {O'Steen}, {Oman}, {Pacifici}, {Pascual}, {Pascual-Granado}, {Patil}, {Perren}, {Pickering}, {Rastogi}, {Roulston}, {Ryan}, {Rykoff}, {Sabater}, {Sakurikar}, {Salgado}, {Sanghi}, {Saunders}, {Savchenko}, {Schwardt}, {Seifert-Eckert}, {Shih}, {Jain}, {Shukla}, {Sick}, {Simpson}, {Singanamalla}, {Singer}, {Singhal}, {Sinha}, {Sip{\H{o}}cz}, {Spitler}, {Stansby}, {Streicher}, {{\v{S}}umak}, {Swinbank}, {Taranu}, {Tewary}, {Tremblay}, {de Val-Borro}, {Van Kooten}, {Vasovi{\'c}}, {Verma}, {de Miranda Cardoso}, {Williams}, {Wilson}, {Winkel}, {Wood-Vasey}, {Xue}, {Yoachim}, {Zhang}, {Zonca}, \& {Astropy Project Contributors}}]{Astropy2022ApJ...935..167A}
{Astropy Collaboration}, {Price-Whelan}, A.~M., {Lim}, P.~L., {et~al.} 2022, \apj, 935, 167, \dodoi{10.3847/1538-4357/ac7c74}

\bibitem[{{Baldwin} {et~al.}(1981){Baldwin}, {Phillips}, \& {Terlevich}}]{BPT1981PASP...93....5B}
{Baldwin}, J.~A., {Phillips}, M.~M., \& {Terlevich}, R. 1981, \pasp, 93, 5, \dodoi{10.1086/130766}

\bibitem[{{Bannister} {et~al.}(2019){Bannister}, {Deller}, {Phillips}, {Macquart}, {Prochaska}, {Tejos}, {Ryder}, {Sadler}, {Shannon}, {Simha}, {Day}, {McQuinn}, {North-Hickey}, {Bhandari}, {Arcus}, {Bennert}, {Burchett}, {Bouwhuis}, {Dodson}, {Ekers}, {Farah}, {Flynn}, {James}, {Kerr}, {Lenc}, {Mahony}, {O'Meara}, {Os{\l}owski}, {Qiu}, {Treu}, {U}, {Bateman}, {Bock}, {Bolton}, {Brown}, {Bunton}, {Chippendale}, {Cooray}, {Cornwell}, {Gupta}, {Hayman}, {Kesteven}, {Koribalski}, {MacLeod}, {McClure-Griffiths}, {Neuhold}, {Norris}, {Pilawa}, {Qiao}, {Reynolds}, {Roxby}, {Shimwell}, {Voronkov}, \& {Wilson}}]{1809242019Sci...365..565B}
{Bannister}, K.~W., {Deller}, A.~T., {Phillips}, C., {et~al.} 2019, Science, 365, 565, \dodoi{10.1126/science.aaw5903}

\bibitem[{{Bell} {et~al.}(2003){Bell}, {McIntosh}, {Katz}, \& {Weinberg}}]{Bell2003ApJS..149..289B}
{Bell}, E.~F., {McIntosh}, D.~H., {Katz}, N., \& {Weinberg}, M.~D. 2003, \apjs, 149, 289, \dodoi{10.1086/378847}

\bibitem[{{Bhandari} {et~al.}(2020{\natexlab{a}}){Bhandari}, {Sadler}, {Prochaska}, {Simha}, {Ryder}, {Marnoch}, {Bannister}, {Macquart}, {Flynn}, {Shannon}, {Tejos}, {Corro-Guerra}, {Day}, {Deller}, {Ekers}, {Lopez}, {Mahony}, {Nu{\~n}ez}, \& {Phillips}}]{Bhandari2020ApJ...895L..37B}
{Bhandari}, S., {Sadler}, E.~M., {Prochaska}, J.~X., {et~al.} 2020{\natexlab{a}}, \apjl, 895, L37, \dodoi{10.3847/2041-8213/ab672e}

\bibitem[{{Bhandari} {et~al.}(2020{\natexlab{b}}){Bhandari}, {Sadler}, {Prochaska}, {Simha}, {Ryder}, {Marnoch}, {Bannister}, {Macquart}, {Flynn}, {Shannon}, {Tejos}, {Corro-Guerra}, {Day}, {Deller}, {Ekers}, {Lopez}, {Mahony}, {Nu{\~n}ez}, \& {Phillips}}]{1906082020ApJ...895L..37B}
---. 2020{\natexlab{b}}, \apjl, 895, L37, \dodoi{10.3847/2041-8213/ab672e}

\bibitem[{{Bhandari} {et~al.}(2022{\natexlab{a}}){Bhandari}, {Gordon}, {Scott}, {Marnoch}, {Sridhar}, {Kumar}, {James}, {Qiu}, {Bannister}, {Deller}, {Eftekhari}, {Fong}, {Glowacki}, {Prochaska}, {Ryder}, {Shannon}, \& {Simha}}]{Bhandari2101172022arXiv221116790B}
{Bhandari}, S., {Gordon}, A.~C., {Scott}, D.~R., {et~al.} 2022{\natexlab{a}}, arXiv e-prints, arXiv:2211.16790, \dodoi{10.48550/arXiv.2211.16790}

\bibitem[{{Bhandari} {et~al.}(2022{\natexlab{b}}){Bhandari}, {Heintz}, {Aggarwal}, {Marnoch}, {Day}, {Sydnor}, {Burke-Spolaor}, {Law}, {Xavier Prochaska}, {Tejos}, {Bannister}, {Butler}, {Deller}, {Ekers}, {Flynn}, {Fong}, {James}, {Lazio}, {Luo}, {Mahony}, {Ryder}, {Sadler}, {Shannon}, {Han}, {Lee}, \& {Zhang}}]{Bhandari2022AJ....163...69B}
{Bhandari}, S., {Heintz}, K.~E., {Aggarwal}, K., {et~al.} 2022{\natexlab{b}}, \aj, 163, 69, \dodoi{10.3847/1538-3881/ac3aec}

\bibitem[{{Bhandari} {et~al.}(2023){Bhandari}, {Marcote}, {Sridhar}, {Eftekhari}, {Hessels}, {Hewitt}, {Kirsten}, {Ould-Boukattine}, {Paragi}, \& {Snelders}}]{190520PRS_VLBI}
{Bhandari}, S., {Marcote}, B., {Sridhar}, N., {et~al.} 2023, arXiv e-prints, arXiv:2308.12801, \dodoi{10.48550/arXiv.2308.12801}

\bibitem[{Bradley {et~al.}(2024)Bradley, Sip{\H o}cz, Robitaille, Tollerud, Vin{\'{\i}}cius, Deil, Barbary, Wilson, Busko, Donath, G{\"u}nther, Cara, Lim, Me{\ss}linger, Burnett, Conseil, Droettboom, Bostroem, Bray, Bratholm, Jamieson, Ginsburg, Barentsen, Craig, Pascual, Rathi, Perrin, Morris, \& Perren}]{larry_bradley_2024_12585239}
Bradley, L., Sip{\H o}cz, B., Robitaille, T., {et~al.} 2024, astropy/photutils: 1.13.0, 1.13.0,  Zenodo, \dodoi{10.5281/zenodo.12585239}

\bibitem[{{Caleb} {et~al.}(2023){Caleb}, {Driessen}, {Gordon}, {Tejos}, {Bernales}, {Qiu}, {Chibueze}, {Stappers}, {Rajwade}, {Cavallaro}, {Wang}, {Kumar}, {Majid}, {Wharton}, {Naudet}, {Bezuidenhout}, {Jankowski}, {Malenta}, {Morello}, {Sanidas}, {Surnis}, {Barr}, {Chen}, {Kramer}, {Fong}, {Kilpatrick}, {Prochaska}, {Simha}, {Venter}, {Heywood}, {Kundu}, \& {Schussler}}]{Caleb2023MNRAS.524.2064C}
{Caleb}, M., {Driessen}, L.~N., {Gordon}, A.~C., {et~al.} 2023, \mnras, 524, 2064, \dodoi{10.1093/mnras/stad1839}

\bibitem[{{Chatterjee} {et~al.}(2017){Chatterjee}, {Law}, {Wharton}, {Burke-Spolaor}, {Hessels}, {Bower}, {Cordes}, {Tendulkar}, {Bassa}, {Demorest}, {Butler}, {Seymour}, {Scholz}, {Abruzzo}, {Bogdanov}, {Kaspi}, {Keimpema}, {Lazio}, {Marcote}, {McLaughlin}, {Paragi}, {Ransom}, {Rupen}, {Spitler}, \& {van Langevelde}}]{Chatterjee121102-2017Natur.541...58C}
{Chatterjee}, S., {Law}, C.~J., {Wharton}, R.~S., {et~al.} 2017, \nat, 541, 58, \dodoi{10.1038/nature20797}

\bibitem[{{Chen} {et~al.}(2025){Chen}, {Tsai}, {Li}, {Wang}, {Feng}, {Zhang}, {Li}, {Zhang}, {Bao}, {Liao}, {Zhang}, {Zuo}, {Bao}, {Niu}, {Luo}, {Zhu}, {Zou}, {Xue}, \& {Zhang}}]{chen2025ApJ...980L..24C}
{Chen}, X.-L., {Tsai}, C.-W., {Li}, D., {et~al.} 2025, \apjl, 980, L24, \dodoi{10.3847/2041-8213/adadfd}

\bibitem[{{CHIME/FRB Collaboration} {et~al.}(2020){CHIME/FRB Collaboration}, {Andersen}, {Bandura}, {Bhardwaj}, {Bij}, {Boyce}, {Boyle}, {Brar}, {Cassanelli}, {Chawla}, {Chen}, {Cliche}, {Cook}, {Cubranic}, {Curtin}, {Denman}, {Dobbs}, {Dong}, {Fandino}, {Fonseca}, {Gaensler}, {Giri}, {Good}, {Halpern}, {Hill}, {Hinshaw}, {H{\"o}fer}, {Josephy}, {Kania}, {Kaspi}, {Landecker}, {Leung}, {Li}, {Lin}, {Masui}, {McKinven}, {Mena-Parra}, {Merryfield}, {Meyers}, {Michilli}, {Milutinovic}, {Mirhosseini}, {M{\"u}nchmeyer}, {Naidu}, {Newburgh}, {Ng}, {Patel}, {Pen}, {Pinsonneault-Marotte}, {Pleunis}, {Quine}, {Rafiei-Ravandi}, {Rahman}, {Ransom}, {Renard}, {Sanghavi}, {Scholz}, {Shaw}, {Shin}, {Siegel}, {Singh}, {Smegal}, {Smith}, {Stairs}, {Tan}, {Tendulkar}, {Tretyakov}, {Vanderlinde}, {Wang}, {Wulf}, \& {Zwaniga}}]{SGR19352020Natur.587...54C}
{CHIME/FRB Collaboration}, {Andersen}, B.~C., {Bandura}, K.~M., {et~al.} 2020, \nat, 587, 54, \dodoi{10.1038/s41586-020-2863-y}

\bibitem[{{Cid Fernandes} {et~al.}(2010){Cid Fernandes}, {Stasi{\'n}ska}, {Schlickmann}, {Mateus}, {Vale Asari}, {Schoenell}, \& {Sodr{\'e}}}]{Fernandes2010MNRAS.403.1036C}
{Cid Fernandes}, R., {Stasi{\'n}ska}, G., {Schlickmann}, M.~S., {et~al.} 2010, \mnras, 403, 1036, \dodoi{10.1111/j.1365-2966.2009.16185.x}

\bibitem[{{Cordes} \& {Lazio}(2002)}]{NE2001-12002astro.ph..7156C}
{Cordes}, J.~M., \& {Lazio}, T.~J.~W. 2002, arXiv e-prints, astro, \dodoi{10.48550/arXiv.astro-ph/0207156}

\bibitem[{{Cordes} \& {Lazio}(2003)}]{NE2001-22003astro.ph..1598C}
---. 2003, arXiv e-prints, astro, \dodoi{10.48550/arXiv.astro-ph/0301598}

\bibitem[{{Cordes} {et~al.}(2016){Cordes}, {Wharton}, {Spitler}, {Chatterjee}, \& {Wasserman}}]{2016arXiv160505890C}
{Cordes}, J.~M., {Wharton}, R.~S., {Spitler}, L.~G., {Chatterjee}, S., \& {Wasserman}, I. 2016, arXiv e-prints, arXiv:1605.05890, \dodoi{10.48550/arXiv.1605.05890}

\bibitem[{Craig {et~al.}(2017)Craig, Crawford, Seifert, Robitaille, Sip{\H o}cz, Walawender, Vin{\'{\i}}cius, Ninan, Droettboom, Youn, Tollerud, Bray, Walker, Janga, Stotts, G{\"u}nther, Rol, Bach, Bradley, Deil, Price-Whelan, Barbary, Horton, Schoenell, Heidt, Gasdia, Nelson, \& Streicher}]{ccdprocmatt_craig_2017_1069648}
Craig, M., Crawford, S., Seifert, M., {et~al.} 2017, astropy/ccdproc: v1.3.0.post1, \dodoi{10.5281/zenodo.1069648}

\bibitem[{{Day} {et~al.}(2020){Day}, {Deller}, {Shannon}, {Qiu(邱昊)}, {Bannister}, {Bhandari}, {Ekers}, {Flynn}, {James}, {Macquart}, {Mahony}, {Phillips}, \& {Xavier Prochaska}}]{1906082020MNRAS.497.3335D}
{Day}, C.~K., {Deller}, A.~T., {Shannon}, R.~M., {et~al.} 2020, \mnras, 497, 3335, \dodoi{10.1093/mnras/staa2138}

\bibitem[{{Dong} {et~al.}(2024){Dong}, {Eftekhari}, {Fong}, {Deller}, {Mannings}, {Simha}, {Sridhar}, {Rafelski}, {Gordon}, {Bhandari}, {Day}, {Heintz}, {Hessels}, {Leja}, {James}, {Kilpatrick}, {Mahony}, {Marcote}, {Margalit}, {Nimmo}, {Prochaska}, {Escorial}, {Ryder}, {Schroeder}, {Shannon}, \& {Tejos}}]{Dong2024ApJ...961...44D}
{Dong}, Y., {Eftekhari}, T., {Fong}, W.-f., {et~al.} 2024, \apj, 961, 44, \dodoi{10.3847/1538-4357/ad0cbd}

\bibitem[{Eftekhari {et~al.}(2019)Eftekhari, Berger, Margalit, Blanchard, Patton, Demorest, Williams, Chatterjee, Cordes, Lunnan, Metzger, \& Nicholl}]{Eftekhari2019}
Eftekhari, T., Berger, E., Margalit, B., {et~al.} 2019, The Astrophysical Journal Letters, 876, L10, \dodoi{10.3847/2041-8213/ab18a5}

\bibitem[{{Feng} {et~al.}(2022{\natexlab{a}}){Feng}, {Zhang}, {Li}, {Yang}, {Wang}, {Niu}, {Dai}, \& {Yao}}]{2022SciBu..67.2398F}
{Feng}, Y., {Zhang}, Y.-K., {Li}, D., {et~al.} 2022{\natexlab{a}}, Science Bulletin, 67, 2398, \dodoi{10.1016/j.scib.2022.11.014}

\bibitem[{{Feng} {et~al.}(2022{\natexlab{b}}){Feng}, {Li}, {Yang}, {Zhang}, {Zhu}, {Zhang}, {Lu}, {Wang}, {Dai}, {Lynch}, {Yao}, {Jiang}, {Niu}, {Zhou}, {Xu}, {Miao}, {Niu}, {Meng}, {Qian}, {Tsai}, {Wang}, {Xue}, {Yue}, {Yuan}, {Zhang}, \& {Zhang}}]{FengsigRM2022Sci...375.1266F}
{Feng}, Y., {Li}, D., {Yang}, Y.-P., {et~al.} 2022{\natexlab{b}}, Science, 375, 1266, \dodoi{10.1126/science.abl7759}

\bibitem[{{Fitzpatrick} \& {Massa}(2007)}]{2007ApJ...663..320F}
{Fitzpatrick}, E.~L., \& {Massa}, D. 2007, \apj, 663, 320, \dodoi{10.1086/518158}

\bibitem[{{Foreman-Mackey} {et~al.}(2013){Foreman-Mackey}, {Hogg}, {Lang}, \& {Goodman}}]{emcmc2013PASP..125..306F}
{Foreman-Mackey}, D., {Hogg}, D.~W., {Lang}, D., \& {Goodman}, J. 2013, \pasp, 125, 306, \dodoi{10.1086/670067}

\bibitem[{{Gordon} {et~al.}(2023){Gordon}, {Fong}, {Kilpatrick}, {Eftekhari}, {Leja}, {Prochaska}, {Nugent}, {Bhandari}, {Blanchard}, {Caleb}, {Day}, {Deller}, {Dong}, {Glowacki}, {Gourdji}, {Mannings}, {Mahoney}, {Marnoch}, {Miller}, {Paterson}, {Rastinejad}, {Ryder}, {Sadler}, {Scott}, {Sears}, {Shannon}, {Simha}, {Stappers}, \& {Tejos}}]{Gordon2023hostgalaxies}
{Gordon}, A.~C., {Fong}, W.-f., {Kilpatrick}, C.~D., {et~al.} 2023, \apj, 954, 80, \dodoi{10.3847/1538-4357/ace5aa}

\bibitem[{{Griffith} {et~al.}(2011){Griffith}, {Tsai}, {Stern}, {Blain}, {Eisenhardt}, {Harrison}, {Jarrett}, {Madsen}, {Stanford}, {Wright}, {Wu}, {Wu}, \& {Yan}}]{Griffith2011ApJ...736L..22G}
{Griffith}, R.~L., {Tsai}, C.-W., {Stern}, D., {et~al.} 2011, \apjl, 736, L22, \dodoi{10.1088/2041-8205/736/1/L22}

\bibitem[{{Haffner} {et~al.}(2009){Haffner}, {Dettmar}, {Beckman}, {Wood}, {Slavin}, {Giammanco}, {Madsen}, {Zurita}, \& {Reynolds}}]{WIM2009RvMP...81..969H}
{Haffner}, L.~M., {Dettmar}, R.~J., {Beckman}, J.~E., {et~al.} 2009, Reviews of Modern Physics, 81, 969, \dodoi{10.1103/RevModPhys.81.969}

\bibitem[{{Hao} {et~al.}(2011){Hao}, {Kennicutt}, {Johnson}, {Calzetti}, {Dale}, \& {Moustakas}}]{HaoCN2011}
{Hao}, C.-N., {Kennicutt}, R.~C., {Johnson}, B.~D., {et~al.} 2011, \apj, 741, 124, \dodoi{10.1088/0004-637X/741/2/124}

\bibitem[{{Hoffmann} {et~al.}(2021){Hoffmann}, {Mack}, {Avila}, {Martlin}, {Cohen}, \& {Bajaj}}]{drizzle2021AAS...23821602H}
{Hoffmann}, S.~L., {Mack}, J., {Avila}, R., {et~al.} 2021, in American Astronomical Society Meeting Abstracts, Vol. 238, American Astronomical Society Meeting Abstracts, 216.02

\bibitem[{{Izotov} {et~al.}(2006){Izotov}, {Stasi{\'n}ska}, {Meynet}, {Guseva}, \& {Thuan}}]{Izotov2006A&A...448..955I}
{Izotov}, Y.~I., {Stasi{\'n}ska}, G., {Meynet}, G., {Guseva}, N.~G., \& {Thuan}, T.~X. 2006, \aap, 448, 955, \dodoi{10.1051/0004-6361:20053763}

\bibitem[{{Karachentsev} \& {Kaisina}(2013)}]{2013AJ....146...46K}
{Karachentsev}, I.~D., \& {Kaisina}, E.~I. 2013, \aj, 146, 46, \dodoi{10.1088/0004-6256/146/3/46}

\bibitem[{{Kauffmann} {et~al.}(2003){Kauffmann}, {Heckman}, {Tremonti}, {Brinchmann}, {Charlot}, {White}, {Ridgway}, {Brinkmann}, {Fukugita}, {Hall}, {Ivezi{\'c}}, {Richards}, \& {Schneider}}]{Kauffmann2003MNRAS.346.1055K}
{Kauffmann}, G., {Heckman}, T.~M., {Tremonti}, C., {et~al.} 2003, \mnras, 346, 1055, \dodoi{10.1111/j.1365-2966.2003.07154.x}

\bibitem[{{Kennicutt} \& {Evans}(2012)}]{SFR2012ARA&A..50..531K}
{Kennicutt}, R.~C., \& {Evans}, N.~J. 2012, \araa, 50, 531, \dodoi{10.1146/annurev-astro-081811-125610}

\bibitem[{{Kewley} {et~al.}(2006){Kewley}, {Groves}, {Kauffmann}, \& {Heckman}}]{Kewley2006MNRAS.372..961K}
{Kewley}, L.~J., {Groves}, B., {Kauffmann}, G., \& {Heckman}, T. 2006, \mnras, 372, 961, \dodoi{10.1111/j.1365-2966.2006.10859.x}

\bibitem[{{Law} {et~al.}(2018){Law}, {Bower}, {Burke-Spolaor}, {Butler}, {Demorest}, {Halle}, {Khudikyan}, {Lazio}, {Pokorny}, {Robnett}, \& {Rupen}}]{realfast}
{Law}, C.~J., {Bower}, G.~C., {Burke-Spolaor}, S., {et~al.} 2018, \apjs, 236, 8, \dodoi{10.3847/1538-4365/aab77b}

\bibitem[{{Law} {et~al.}(2024){Law}, {Sharma}, {Ravi}, {Chen}, {Catha}, {Connor}, {Faber}, {Hallinan}, {Harnach}, {Hellbourg}, {Hobbs}, {Hodge}, {Hodges}, {Lamb}, {Rasmussen}, {Sherman}, {Shi}, {Simard}, {Squillace}, {Weinreb}, {Woody}, \& {Yurk}}]{LawDSA2024ApJ...967...29L}
{Law}, C.~J., {Sharma}, K., {Ravi}, V., {et~al.} 2024, \apj, 967, 29, \dodoi{10.3847/1538-4357/ad3736}

\bibitem[{{Lee} {et~al.}(2023){Lee}, {Khrykin}, {Simha}, {Ata}, {Huang}, {Prochaska}, {Tejos}, {Cooke}, {Nagamine}, \& {Zhang}}]{foreground2023ApJ...954L...7L}
{Lee}, K.-G., {Khrykin}, I.~S., {Simha}, S., {et~al.} 2023, \apjl, 954, L7, \dodoi{10.3847/2041-8213/acefb5}

\bibitem[{Li {et~al.}(2018)Li, Wang, Qian, Krco, Dunning, Jiang, Yue, Jin, Zhu, Pan, \& Nan}]{LiCRAFTS2018}
Li, D., Wang, P., Qian, L., {et~al.} 2018, IEEE Microwave Magazine, 19, 112, \dodoi{10.1109/MMM.2018.2802178}

\bibitem[{{Li} {et~al.}(2023){Li}, {Tsai}, {Stern}, {Wu}, {Assef}, {Blain}, {D{\'\i}az-Santos}, {Eisenhardt}, {Griffith}, {Jarrett}, {Jun}, {Lake}, \& {Saade}}]{2023ApJ...958..162L}
{Li}, G., {Tsai}, C.-W., {Stern}, D., {et~al.} 2023, \apj, 958, 162, \dodoi{10.3847/1538-4357/ace25b}

\bibitem[{{Li} {et~al.}(2020){Li}, {Yang}, \& {Dai}}]{LYD2020ApJ...896...71L}
{Li}, Q.-C., {Yang}, Y.-P., \& {Dai}, Z.-G. 2020, \apj, 896, 71, \dodoi{10.3847/1538-4357/ab8db8}

\bibitem[{{Lorimer} {et~al.}(2007){Lorimer}, {Bailes}, {McLaughlin}, {Narkevic}, \& {Crawford}}]{Lorimer2007Sci...318..777L}
{Lorimer}, D.~R., {Bailes}, M., {McLaughlin}, M.~A., {Narkevic}, D.~J., \& {Crawford}, F. 2007, Science, 318, 777, \dodoi{10.1126/science.1147532}

\bibitem[{{Luo} {et~al.}(2020){Luo}, {Wang}, {Men}, {Zhang}, {Jiang}, {Xu}, {Wang}, {Lee}, {Han}, {Zhang}, {Caballero}, {Chen}, {Chen}, {Gan}, {Guo}, {Hao}, {Huang}, {Jiang}, {Li}, {Li}, {Li}, {Luo}, {Pan}, {Pei}, {Qian}, {Sun}, {Wang}, {Wang}, {Wen}, {Xu}, {Xu}, {Yan}, {Yan}, {Yu}, {Yuan}, {Zhang}, \& {Zhu}}]{Luo1803012020Natur.586..693L}
{Luo}, R., {Wang}, B.~J., {Men}, Y.~P., {et~al.} 2020, \nat, 586, 693, \dodoi{10.1038/s41586-020-2827-2}

\bibitem[{{Macquart} {et~al.}(2020){Macquart}, {Prochaska}, {McQuinn}, {Bannister}, {Bhandari}, {Day}, {Deller}, {Ekers}, {James}, {Marnoch}, {Os{\l}owski}, {Phillips}, {Ryder}, {Scott}, {Shannon}, \& {Tejos}}]{MacquartRelation2020Natur.581..391M}
{Macquart}, J.~P., {Prochaska}, J.~X., {McQuinn}, M., {et~al.} 2020, \nat, 581, 391, \dodoi{10.1038/s41586-020-2300-2}

\bibitem[{{Maguire} {et~al.}(2018){Maguire}, {Sim}, {Shingles}, {Spyromilio}, {Jerkstrand}, {Sullivan}, {Chen}, {Cartier}, {Dimitriadis}, {Frohmaier}, {Galbany}, {Guti{\'e}rrez}, {Hosseinzadeh}, {Howell}, {Inserra}, {Rudy}, \& {Sollerman}}]{IaSpec2018MNRAS.477.3567M}
{Maguire}, K., {Sim}, S.~A., {Shingles}, L., {et~al.} 2018, \mnras, 477, 3567, \dodoi{10.1093/mnras/sty820}

\bibitem[{{Marcote} {et~al.}(2020){Marcote}, {Nimmo}, {Hessels}, {Tendulkar}, {Bassa}, {Paragi}, {Keimpema}, {Bhardwaj}, {Karuppusamy}, {Kaspi}, {Law}, {Michilli}, {Aggarwal}, {Andersen}, {Archibald}, {Bandura}, {Bower}, {Boyle}, {Brar}, {Burke-Spolaor}, {Butler}, {Cassanelli}, {Chawla}, {Demorest}, {Dobbs}, {Fonseca}, {Giri}, {Good}, {Gourdji}, {Josephy}, {Kirichenko}, {Kirsten}, {Landecker}, {Lang}, {Lazio}, {Li}, {Lin}, {Linford}, {Masui}, {Mena-Parra}, {Naidu}, {Ng}, {Patel}, {Pen}, {Pleunis}, {Rafiei-Ravandi}, {Rahman}, {Renard}, {Scholz}, {Siegel}, {Smith}, {Stairs}, {Vanderlinde}, \& {Zwaniga}}]{1809162020Natur.577..190M}
{Marcote}, B., {Nimmo}, K., {Hessels}, J.~W.~T., {et~al.} 2020, \nat, 577, 190, \dodoi{10.1038/s41586-019-1866-z}

\bibitem[{{Michilli} {et~al.}(2018){Michilli}, {Seymour}, {Hessels}, {Spitler}, {Gajjar}, {Archibald}, {Bower}, {Chatterjee}, {Cordes}, {Gourdji}, {Heald}, {Kaspi}, {Law}, {Sobey}, {Adams}, {Bassa}, {Bogdanov}, {Brinkman}, {Demorest}, {Fernandez}, {Hellbourg}, {Lazio}, {Lynch}, {Maddox}, {Marcote}, {McLaughlin}, {Paragi}, {Ransom}, {Scholz}, {Siemion}, {Tendulkar}, {van Rooy}, {Wharton}, \& {Whitlow}}]{Michilli201102-2018Natur.553..182M}
{Michilli}, D., {Seymour}, A., {Hessels}, J.~W.~T., {et~al.} 2018, \nat, 553, 182, \dodoi{10.1038/nature25149}

\bibitem[{{Moustakas} {et~al.}(2013){Moustakas}, {Coil}, {Aird}, {Blanton}, {Cool}, {Eisenstein}, {Mendez}, {Wong}, {Zhu}, \& {Arnouts}}]{Moustakas2013ApJ...767...50M}
{Moustakas}, J., {Coil}, A.~L., {Aird}, J., {et~al.} 2013, \apj, 767, 50, \dodoi{10.1088/0004-637X/767/1/50}

\bibitem[{{Murphy} {et~al.}(2011){Murphy}, {Condon}, {Schinnerer}, {Kennicutt}, {Calzetti}, {Armus}, {Helou}, {Turner}, {Aniano}, {Beir{\~a}o}, {Bolatto}, {Brandl}, {Croxall}, {Dale}, {Donovan Meyer}, {Draine}, {Engelbracht}, {Hunt}, {Hao}, {Koda}, {Roussel}, {Skibba}, \& {Smith}}]{MurphyEJ2011}
{Murphy}, E.~J., {Condon}, J.~J., {Schinnerer}, E., {et~al.} 2011, \apj, 737, 67, \dodoi{10.1088/0004-637X/737/2/67}

\bibitem[{{Nan} {et~al.}(2011){Nan}, {Li}, {Jin}, {Wang}, {Zhu}, {Zhu}, {Zhang}, {Yue}, \& {Qian}}]{NanFAST2011IJMPD..20..989N}
{Nan}, R., {Li}, D., {Jin}, C., {et~al.} 2011, International Journal of Modern Physics D, 20, 989, \dodoi{10.1142/S0218271811019335}

\bibitem[{{Nicholl} {et~al.}(2013){Nicholl}, {Smartt}, {Jerkstrand}, {Inserra}, {McCrum}, {Kotak}, {Fraser}, {Wright}, {Chen}, {Smith}, {Young}, {Sim}, {Valenti}, {Howell}, {Bresolin}, {Kudritzki}, {Tonry}, {Huber}, {Rest}, {Pastorello}, {Tomasella}, {Cappellaro}, {Benetti}, {Mattila}, {Kankare}, {Kangas}, {Leloudas}, {Sollerman}, {Taddia}, {Berger}, {Chornock}, {Narayan}, {Stubbs}, {Foley}, {Lunnan}, {Soderberg}, {Sanders}, {Milisavljevic}, {Margutti}, {Kirshner}, {Elias-Rosa}, {Morales-Garoffolo}, {Taubenberger}, {Botticella}, {Gezari}, {Urata}, {Rodney}, {Riess}, {Scolnic}, {Wood-Vasey}, {Burgett}, {Chambers}, {Flewelling}, {Magnier}, {Kaiser}, {Metcalfe}, {Morgan}, {Price}, {Sweeney}, \& {Waters}}]{SLSNe2013Natur.502..346N}
{Nicholl}, M., {Smartt}, S.~J., {Jerkstrand}, A., {et~al.} 2013, \nat, 502, 346, \dodoi{10.1038/nature12569}

\bibitem[{{Niu} {et~al.}(2022){Niu}, {Aggarwal}, {Li}, {Zhang}, {Chatterjee}, {Tsai}, {Yu}, {Law}, {Burke-Spolaor}, {Cordes}, {Zhang}, {Ocker}, {Yao}, {Wang}, {Feng}, {Niino}, {Bochenek}, {Cruces}, {Connor}, {Jiang}, {Dai}, {Luo}, {Li}, {Miao}, {Niu}, {Anna-Thomas}, {Sydnor}, {Stern}, {Wang}, {Yuan}, {Yue}, {Zhou}, {Yan}, {Zhu}, \& {Zhang}}]{Niu190520-2022Natur.606..873N}
{Niu}, C.~H., {Aggarwal}, K., {Li}, D., {et~al.} 2022, \nat, 606, 873, \dodoi{10.1038/s41586-022-04755-5}

\bibitem[{{Nugent} {et~al.}(2022){Nugent}, {Fong}, {Dong}, {Leja}, {Berger}, {Zevin}, {Chornock}, {Cobb}, {Kelley}, {Kilpatrick}, {Levan}, {Margutti}, {Paterson}, {Perley}, {Escorial}, {Smith}, \& {Tanvir}}]{Nugent2022ApJ...940...57N}
{Nugent}, A.~E., {Fong}, W.-F., {Dong}, Y., {et~al.} 2022, \apj, 940, 57, \dodoi{10.3847/1538-4357/ac91d1}

\bibitem[{{Ocker} {et~al.}(2023){Ocker}, {Cordes}, {Chatterjee}, {Li}, {Niu}, {McKee}, {Law}, \& {Anna-Thomas}}]{OckerScatter2023MNRAS.519..821O}
{Ocker}, S.~K., {Cordes}, J.~M., {Chatterjee}, S., {et~al.} 2023, \mnras, 519, 821, \dodoi{10.1093/mnras/stac3547}

\bibitem[{{Ocker} {et~al.}(2022){Ocker}, {Cordes}, {Chatterjee}, {Niu}, {Li}, {McKee}, {Law}, {Tsai}, {Anna-Thomas}, {Yao}, \& {Cruces}}]{OckerHost2022ApJ...931...87O}
---. 2022, \apj, 931, 87, \dodoi{10.3847/1538-4357/ac6504}

\bibitem[{{Petroff} {et~al.}(2022){Petroff}, {Hessels}, \& {Lorimer}}]{Petroff2022A&ARv..30....2P}
{Petroff}, E., {Hessels}, J.~W.~T., \& {Lorimer}, D.~R. 2022, \aapr, 30, 2, \dodoi{10.1007/s00159-022-00139-w}

\bibitem[{{Planck Collaboration} {et~al.}(2020){Planck Collaboration}, {Aghanim}, {Akrami}, {Ashdown}, {Aumont}, {Baccigalupi}, {Ballardini}, {Banday}, {Barreiro}, {Bartolo}, {Basak}, {Battye}, {Benabed}, {Bernard}, {Bersanelli}, {Bielewicz}, {Bock}, {Bond}, {Borrill}, {Bouchet}, {Boulanger}, {Bucher}, {Burigana}, {Butler}, {Calabrese}, {Cardoso}, {Carron}, {Challinor}, {Chiang}, {Chluba}, {Colombo}, {Combet}, {Contreras}, {Crill}, {Cuttaia}, {de Bernardis}, {de Zotti}, {Delabrouille}, {Delouis}, {Di Valentino}, {Diego}, {Dor{\'e}}, {Douspis}, {Ducout}, {Dupac}, {Dusini}, {Efstathiou}, {Elsner}, {En{\ss}lin}, {Eriksen}, {Fantaye}, {Farhang}, {Fergusson}, {Fernandez-Cobos}, {Finelli}, {Forastieri}, {Frailis}, {Fraisse}, {Franceschi}, {Frolov}, {Galeotta}, {Galli}, {Ganga}, {G{\'e}nova-Santos}, {Gerbino}, {Ghosh}, {Gonz{\'a}lez-Nuevo}, {G{\'o}rski}, {Gratton}, {Gruppuso}, {Gudmundsson}, {Hamann}, {Handley}, {Hansen}, {Herranz}, {Hildebrandt}, {Hivon}, {Huang}, {Jaffe}, {Jones}, {Karakci}, {Keih{\"a}nen},
  {Keskitalo}, {Kiiveri}, {Kim}, {Kisner}, {Knox}, {Krachmalnicoff}, {Kunz}, {Kurki-Suonio}, {Lagache}, {Lamarre}, {Lasenby}, {Lattanzi}, {Lawrence}, {Le Jeune}, {Lemos}, {Lesgourgues}, {Levrier}, {Lewis}, {Liguori}, {Lilje}, {Lilley}, {Lindholm}, {L{\'o}pez-Caniego}, {Lubin}, {Ma}, {Mac{\'\i}as-P{\'e}rez}, {Maggio}, {Maino}, {Mandolesi}, {Mangilli}, {Marcos-Caballero}, {Maris}, {Martin}, {Martinelli}, {Mart{\'\i}nez-Gonz{\'a}lez}, {Matarrese}, {Mauri}, {McEwen}, {Meinhold}, {Melchiorri}, {Mennella}, {Migliaccio}, {Millea}, {Mitra}, {Miville-Desch{\^e}nes}, {Molinari}, {Montier}, {Morgante}, {Moss}, {Natoli}, {N{\o}rgaard-Nielsen}, {Pagano}, {Paoletti}, {Partridge}, {Patanchon}, {Peiris}, {Perrotta}, {Pettorino}, {Piacentini}, {Polastri}, {Polenta}, {Puget}, {Rachen}, {Reinecke}, {Remazeilles}, {Renzi}, {Rocha}, {Rosset}, {Roudier}, {Rubi{\~n}o-Mart{\'\i}n}, {Ruiz-Granados}, {Salvati}, {Sandri}, {Savelainen}, {Scott}, {Shellard}, {Sirignano}, {Sirri}, {Spencer}, {Sunyaev}, {Suur-Uski}, {Tauber}, {Tavagnacco},
  {Tenti}, {Toffolatti}, {Tomasi}, {Trombetti}, {Valenziano}, {Valiviita}, {Van Tent}, {Vibert}, {Vielva}, {Villa}, {Vittorio}, {Wandelt}, {Wehus}, {White}, {White}, {Zacchei}, \& {Zonca}}]{Planck2020A&A...641A...6P}
{Planck Collaboration}, {Aghanim}, N., {Akrami}, Y., {et~al.} 2020, \aap, 641, A6, \dodoi{10.1051/0004-6361/201833910}

\bibitem[{{Prochaska} {et~al.}(2019){Prochaska}, {Macquart}, {McQuinn}, {Simha}, {Shannon}, {Day}, {Marnoch}, {Ryder}, {Deller}, {Bannister}, {Bhandari}, {Bordoloi}, {Bunton}, {Cho}, {Flynn}, {Mahony}, {Phillips}, {Qiu}, \& {Tejos}}]{1811122019Sci...366..231P}
{Prochaska}, J.~X., {Macquart}, J.-P., {McQuinn}, M., {et~al.} 2019, Science, 366, 231, \dodoi{10.1126/science.aay0073}

\bibitem[{{Ravi} {et~al.}(2019){Ravi}, {Battaglia}, {Burke-Spolaor}, {Chatterjee}, {Cordes}, {Hallinan}, {Law}, {Lazio}, {Masui}, {McQuinn}, {Mu{\~n}oz}, {Palliyaguru}, {Prochaska}, {Seymour}, {Vedantham}, \& {Zheng}}]{RaviWP2019BAAS...51c.420R}
{Ravi}, V., {Battaglia}, N., {Burke-Spolaor}, S., {et~al.} 2019, \baas, 51, 420, \dodoi{10.48550/arXiv.1903.06535}

\bibitem[{{Ravi} {et~al.}(2022{\natexlab{a}}){Ravi}, {Law}, {Li}, {Aggarwal}, {Bhardwaj}, {Burke-Spolaor}, {Connor}, {Lazio}, {Simard}, {Somalwar}, \& {Tendulkar}}]{Ravi201124A2022MNRAS.513..982R}
{Ravi}, V., {Law}, C.~J., {Li}, D., {et~al.} 2022{\natexlab{a}}, \mnras, 513, 982, \dodoi{10.1093/mnras/stac465}

\bibitem[{{Ravi} {et~al.}(2022{\natexlab{b}}){Ravi}, {Law}, {Li}, {Aggarwal}, {Bhardwaj}, {Burke-Spolaor}, {Connor}, {Lazio}, {Simard}, {Somalwar}, \& {Tendulkar}}]{201124Ravi2022MNRAS.513..982R}
---. 2022{\natexlab{b}}, \mnras, 513, 982, \dodoi{10.1093/mnras/stac465}

\bibitem[{{Ravi} {et~al.}(2023){Ravi}, {Catha}, {Chen}, {Connor}, {Faber}, {Lamb}, {Hallinan}, {Harnach}, {Hellbourg}, {Hobbs}, {Hodge}, {Hodges}, {Law}, {Rasmussen}, {Sharma}, {Sherman}, {Shi}, {Simard}, {Squillace}, {Weinreb}, {Woody}, {Yadlapalli}, {Ahumada}, {Dong}, {Fremling}, {Huang}, {Karambelkar}, \& {Miller}}]{220912Ravi2023ApJ...949L...3R}
{Ravi}, V., {Catha}, M., {Chen}, G., {et~al.} 2023, \apjl, 949, L3, \dodoi{10.3847/2041-8213/acc4b6}

\bibitem[{{Reynolds}(1977)}]{1977ApJ...216..433R}
{Reynolds}, R.~J. 1977, \apj, 216, 433, \dodoi{10.1086/155484}

\bibitem[{{Schlegel} {et~al.}(1998){Schlegel}, {Finkbeiner}, \& {Davis}}]{SFD1998ApJ...500..525S}
{Schlegel}, D.~J., {Finkbeiner}, D.~P., \& {Davis}, M. 1998, \apj, 500, 525, \dodoi{10.1086/305772}

\bibitem[{{Schulze} {et~al.}(2021){Schulze}, {Yaron}, {Sollerman}, {Leloudas}, {Gal}, {Wright}, {Lunnan}, {Gal-Yam}, {Ofek}, {Perley}, {Filippenko}, {Kasliwal}, {Kulkarni}, {Neill}, {Nugent}, {Quimby}, {Sullivan}, {Strotjohann}, {Arcavi}, {Ben-Ami}, {Bianco}, {Bloom}, {De}, {Fraser}, {Fremling}, {Horesh}, {Johansson}, {Kelly}, {Kne{\v{z}}evi{\'c}}, {Kne{\v{z}}evi{\'c}}, {Maguire}, {Nyholm}, {Papadogiannakis}, {Petrushevska}, {Rubin}, {Yan}, {Yang}, {Adams}, {Bufano}, {Clubb}, {Foley}, {Green}, {Harmanen}, {Ho}, {Hook}, {Hosseinzadeh}, {Howell}, {Kong}, {Kotak}, {Matheson}, {McCully}, {Milisavljevic}, {Pan}, {Poznanski}, {Shivvers}, {van Velzen}, \& {Verbeek}}]{CCSNe2021ApJS..255...29S}
{Schulze}, S., {Yaron}, O., {Sollerman}, J., {et~al.} 2021, \apjs, 255, 29, \dodoi{10.3847/1538-4365/abff5e}

\bibitem[{{Sharma} {et~al.}(2024){Sharma}, {Ravi}, {Connor}, {Law}, {Ocker}, {Sherman}, {Kosogorov}, {Faber}, {Hallinan}, {Harnach}, {Hellbourg}, {Hobbs}, {Hodge}, {Hodges}, {Lamb}, {Rasmussen}, {Somalwar}, {Weinreb}, {Woody}, {Leja}, {Anand}, {Das}, {Qin}, {Rose}, {Dong}, {Miller}, \& {Yao}}]{Sharma2024Natur.635...61S}
{Sharma}, K., {Ravi}, V., {Connor}, L., {et~al.} 2024, \nat, 635, 61, \dodoi{10.1038/s41586-024-08074-9}

\bibitem[{{Spitler} {et~al.}(2014){Spitler}, {Cordes}, {Hessels}, {Lorimer}, {McLaughlin}, {Chatterjee}, {Crawford}, {Deneva}, {Kaspi}, {Wharton}, {Allen}, {Bogdanov}, {Brazier}, {Camilo}, {Freire}, {Jenet}, {Karako-Argaman}, {Knispel}, {Lazarus}, {Lee}, {van Leeuwen}, {Lynch}, {Ransom}, {Scholz}, {Siemens}, {Stairs}, {Stovall}, {Swiggum}, {Venkataraman}, {Zhu}, {Aulbert}, \& {Fehrmann}}]{Spitler2014ApJ...790..101S}
{Spitler}, L.~G., {Cordes}, J.~M., {Hessels}, J.~W.~T., {et~al.} 2014, \apj, 790, 101, \dodoi{10.1088/0004-637X/790/2/101}

\bibitem[{{Spitler} {et~al.}(2016){Spitler}, {Scholz}, {Hessels}, {Bogdanov}, {Brazier}, {Camilo}, {Chatterjee}, {Cordes}, {Crawford}, {Deneva}, {Ferdman}, {Freire}, {Kaspi}, {Lazarus}, {Lynch}, {Madsen}, {McLaughlin}, {Patel}, {Ransom}, {Seymour}, {Stairs}, {Stappers}, {van Leeuwen}, \& {Zhu}}]{Spitler2016Natur.531..202S}
{Spitler}, L.~G., {Scholz}, P., {Hessels}, J.~W.~T., {et~al.} 2016, \nat, 531, 202, \dodoi{10.1038/nature17168}

\bibitem[{{Taggart} \& {Perley}(2021)}]{Taggart2021MNRAS.503.3931T}
{Taggart}, K., \& {Perley}, D.~A. 2021, \mnras, 503, 3931, \dodoi{10.1093/mnras/stab174}

\bibitem[{{Tendulkar} {et~al.}(2017){Tendulkar}, {Bassa}, {Cordes}, {Bower}, {Law}, {Chatterjee}, {Adams}, {Bogdanov}, {Burke-Spolaor}, {Butler}, {Demorest}, {Hessels}, {Kaspi}, {Lazio}, {Maddox}, {Marcote}, {McLaughlin}, {Paragi}, {Ransom}, {Scholz}, {Seymour}, {Spitler}, {van Langevelde}, \& {Wharton}}]{Tendulkar121102-2017ApJ...834L...7T}
{Tendulkar}, S.~P., {Bassa}, C.~G., {Cordes}, J.~M., {et~al.} 2017, \apjl, 834, L7, \dodoi{10.3847/2041-8213/834/2/L7}

\bibitem[{{Tian} {et~al.}(2024){Tian}, {Rajwade}, {Pastor-Marazuela}, {Stappers}, {Bezuidenhout}, {Caleb}, {Jankowski}, {Barr}, \& {Kramer}}]{Tian2024MNRAS.533.3174T}
{Tian}, J., {Rajwade}, K.~M., {Pastor-Marazuela}, I., {et~al.} 2024, \mnras, 533, 3174, \dodoi{10.1093/mnras/stae2013}

\bibitem[{{Tody}(1986)}]{IRAF1986SPIE..627..733T}
{Tody}, D. 1986, in Society of Photo-Optical Instrumentation Engineers (SPIE) Conference Series, Vol. 627, Instrumentation in astronomy VI, ed. D.~L. {Crawford}, 733, \dodoi{10.1117/12.968154}

\bibitem[{{Tody}(1993)}]{IRAF1993ASPC...52..173T}
{Tody}, D. 1993, in Astronomical Society of the Pacific Conference Series, Vol.~52, Astronomical Data Analysis Software and Systems II, ed. R.~J. {Hanisch}, R.~J.~V. {Brissenden}, \& J.~{Barnes}, 173

\bibitem[{Virtanen {et~al.}(2020)Virtanen, Gommers, Oliphant, Haberland, Reddy, Cournapeau, Burovski, Peterson, Weckesser, Bright, {van der Walt}, Brett, Wilson, Millman, Mayorov, Nelson, Jones, Kern, Larson, Carey, Polat, Feng, Moore, {VanderPlas}, Laxalde, Perktold, Cimrman, Henriksen, Quintero, Harris, Archibald, Ribeiro, Pedregosa, {van Mulbregt}, \& {SciPy 1.0 Contributors}}]{2020SciPy-NMeth}
Virtanen, P., Gommers, R., Oliphant, T.~E., {et~al.} 2020, Nature Methods, 17, 261, \dodoi{10.1038/s41592-019-0686-2}

\bibitem[{{Wright} {et~al.}(2010){Wright}, {Eisenhardt}, {Mainzer}, {Ressler}, {Cutri}, {Jarrett}, {Kirkpatrick}, {Padgett}, {McMillan}, {Skrutskie}, {Stanford}, {Cohen}, {Walker}, {Mather}, {Leisawitz}, {Gautier}, {McLean}, {Benford}, {Lonsdale}, {Blain}, {Mendez}, {Irace}, {Duval}, {Liu}, {Royer}, {Heinrichsen}, {Howard}, {Shannon}, {Kendall}, {Walsh}, {Larsen}, {Cardon}, {Schick}, {Schwalm}, {Abid}, {Fabinsky}, {Naes}, \& {Tsai}}]{WISE2010AJ....140.1868W}
{Wright}, E.~L., {Eisenhardt}, P. R.~M., {Mainzer}, A.~K., {et~al.} 2010, \aj, 140, 1868, \dodoi{10.1088/0004-6256/140/6/1868}

\bibitem[{{Xu} {et~al.}(2022){Xu}, {Niu}, {Chen}, {Lee}, {Zhu}, {Dong}, {Zhang}, {Jiang}, {Wang}, {Xu}, {Zhang}, {Fu}, {Filippenko}, {Peng}, {Zhou}, {Zhang}, {Wang}, {Feng}, {Li}, {Brink}, {Li}, {Lu}, {Yang}, {Caballero}, {Cai}, {Chen}, {Dai}, {Djorgovski}, {Esamdin}, {Gan}, {Guhathakurta}, {Han}, {Hao}, {Huang}, {Jiang}, {Li}, {Li}, {Li}, {Li}, {Li}, {Liu}, {Luo}, {Men}, {Niu}, {Peng}, {Qian}, {Song}, {Stern}, {Stockton}, {Sun}, {Wang}, {Wang}, {Wang}, {Wang}, {Wu}, {Xiao}, {Xiong}, {Xu}, {Xu}, {Yang}, {Yang}, {Yao}, {Yi}, {Yue}, {Yu}, {Yu}, {Yuan}, {Zhang}, {Zhang}, {Zhang}, {Zhao}, {Zheng}, {Zhu}, \& {Zou}}]{Xu2011242022Natur.609..685X}
{Xu}, H., {Niu}, J.~R., {Chen}, P., {et~al.} 2022, \nat, 609, 685, \dodoi{10.1038/s41586-022-05071-8}

\bibitem[{{Xu} {et~al.}(2023){Xu}, {Feng}, {Li}, {Wang}, {Zhang}, {Xie}, {Chen}, {Wang}, {Kang}, {Hu}, {Zheng}, {Tsai}, {Chen}, \& {Zhou}}]{blinkverse}
{Xu}, J., {Feng}, Y., {Li}, D., {et~al.} 2023, Universe, 9, 330, \dodoi{10.3390/universe9070330}

\bibitem[{{Yamasaki} \& {Totani}(2020)}]{DMhalo2020ApJ...888..105Y}
{Yamasaki}, S., \& {Totani}, T. 2020, \apj, 888, 105, \dodoi{10.3847/1538-4357/ab58c4}

\bibitem[{{Yang} {et~al.}(2024){Yang}, {Feng}, {Tsai}, {Li}, {Shi}, {Wang}, {Yang}, {Zhang}, {Niu}, {Yao}, {Cui}, {Su}, {Li}, {Zhang}, {Zhu}, \& {Cotton}}]{2024ApJ...976..165Y}
{Yang}, A.~Y., {Feng}, Y., {Tsai}, C.-W., {et~al.} 2024, \apj, 976, 165, \dodoi{10.3847/1538-4357/ad7d02}

\bibitem[{{Yang} \& {Dai}(2019)}]{YD2019ApJ...885..149Y}
{Yang}, Y.-H., \& {Dai}, Z.-G. 2019, \apj, 885, 149, \dodoi{10.3847/1538-4357/ab48dd}

\bibitem[{{Yao} {et~al.}(2017){Yao}, {Manchester}, \& {Wang}}]{YMW2017ApJ...835...29Y}
{Yao}, J.~M., {Manchester}, R.~N., \& {Wang}, N. 2017, \apj, 835, 29, \dodoi{10.3847/1538-4357/835/1/29}

\bibitem[{{Zhang}(2018)}]{ZhangDMz2018ApJ...867L..21Z}
{Zhang}, B. 2018, \apjl, 867, L21, \dodoi{10.3847/2041-8213/aae8e3}

\bibitem[{{Zhang}(2023)}]{Zhang2023RvMP...95c5005Z}
---. 2023, Reviews of Modern Physics, 95, 035005, \dodoi{10.1103/RevModPhys.95.035005}

\bibitem[{{Zhang} {et~al.}(2023){Zhang}, {Li}, {Zhang}, {Cao}, {Feng}, {Wang}, {Qu}, {Niu}, {Zhu}, {Han}, {Jiang}, {Lee}, {Li}, {Luo}, {Niu}, {Tsai}, {Wang}, {Wang}, {Wu}, {Xu}, {Yang}, {Zhang}, {Zhou}, \& {Zhu}}]{220912ZYK2023ApJ...955..142Z}
{Zhang}, Y.-K., {Li}, D., {Zhang}, B., {et~al.} 2023, \apj, 955, 142, \dodoi{10.3847/1538-4357/aced0b}

\end{thebibliography}
